\documentclass[twocolumn]{aastex62}
\usepackage{amssymb}
\usepackage{amsmath}
\usepackage{latexsym}
\usepackage{fnpos}
\usepackage{scrextend}
\usepackage{scalerel}
\usepackage{hyperref}
\usepackage{tabularx}
\usepackage{xspace}
\usepackage{enumerate}
\usepackage{graphicx}
\usepackage{url}
\usepackage{longtable}
\usepackage{lipsum}
\usepackage{rotating}
\usepackage{natbib}
\usepackage[maxfloats=256]{morefloats}
\newcommand{\eal}[2]{\ifmmode{\mathrm{#1\,#2}}\else{#1\textsc{$\,$\lowercase{#2}}}\fi\xspace}
\newcommand{\feal}[2]{\ifmmode{\mathrm{#1\,#2}}\else{[#1\textsc{$\,$\lowercase{#2}}]}\fi\xspace}
\newcommand{\hfeal}[2]{\ifmmode{\mathrm{#1\,#2}}\else{#1\textsc{$\,$\lowercase{#2}}]}\fi\xspace}

\graphicspath{{./}{figures/}}

\received{***}
\revised{***}
\accepted{***}
\submitjournal{ApJ}

\shorttitle{X-rays from $\gamma$-ray-detected novae}
\shortauthors{Gordon et al.}

\begin{document}

\title{Surveying the X-ray Behavior of Novae as They Emit $\gamma$-rays}

\correspondingauthor{Alexa Gordon}
\email{muethela@msu.edu, aydielia@pa.msu.edu}

\author[0000-0002-5025-4645]{A.~C.\ Gordon}
\affiliation{Center for Data Intensive and Time Domain Astronomy, Department of Physics and Astronomy, Michigan State University, East Lansing, MI 48824, USA \\}
\affiliation{Center for Interdisciplinary Exploration and Research in Astrophysics and Department of Physics and Astronomy, Northwestern University, 2145 Sheridan Road, Evanston, IL 60208-3112, USA \\}

\author[0000-0001-8525-3442]{E.\ Aydi}
\affil{Center for Data Intensive and Time Domain Astronomy, Department of Physics and Astronomy, Michigan State University, East Lansing, MI 48824, USA \\}

\author[0000-0001-5624-2613]{K.~L.\ Page}
\affiliation{School of Physics and Astronomy, University of Leicester, University Road, Leicester, LE1 7RH, UK\\}

\author{Kwan-Lok Li}
\affiliation{Department of Physics, National Cheng Kung University, 70101 Tainan, Taiwan\\}

\author[0000-0002-8400-3705]{L.\ Chomiuk}
\affiliation{Center for Data Intensive and Time Domain Astronomy, Department of Physics and Astronomy, Michigan State University, East Lansing, MI 48824, USA \\}

\author{K.~V.\ Sokolovsky}
\affiliation{Center for Data Intensive and Time Domain Astronomy, Department of Physics and Astronomy, Michigan State University, East Lansing, MI 48824, USA \\}

\author{K.\ Mukai}
\affiliation{Center for Space Science and Technology, University of Maryland Baltimore County, Baltimore, MD 21250, USA}
\affiliation{CRESST and X-ray Astrophysics Laboratory, NASA/GSFC, Greenbelt, MD 20771, USA\\}

\author{J.\ Seitz}
\affiliation{Center for Data Intensive and Time Domain Astronomy, Department of Physics and Astronomy, Michigan State University, East Lansing, MI 48824, USA \\}


\begin{abstract}
The detection of GeV $\gamma$-ray emission from Galactic novae by \emph{Fermi}-LAT has become routine since 2010, and is generally associated with shocks internal to the nova ejecta.
These
shocks are also expected to heat plasma to $\sim 10^7$\,K, resulting in detectable X-ray emission. In this paper, we investigate 13 $\gamma$-ray emitting novae observed with the \emph{Neil Gehrels Swift Observatory}, searching for 1--10 keV X-ray emission concurrent with $\gamma$-ray detections. We also analyze $\gamma$-ray observations of novae V407~Lup (2016) and V357~Mus (2018). We find that most novae do eventually show X-ray evidence of hot shocked plasma, but not until the $\gamma$-rays have faded below detectability. We suggest that the delayed rise of the X-ray emission is due to large absorbing columns and/or X-ray suppression by corrugated shock fronts. The only nova in our sample with a concurrent X-ray/$\gamma$-ray detection is also the only embedded nova (V407~Cyg). 
This exception supports a scenario where novae with giant companions produce shocks with external circumbinary material and are characterized by lower density environments, in comparison with novae with dwarf companions where shocks occur internal to the dense ejecta. 
\end{abstract}


\section{Introduction} \label{sec:intro}
A classical nova is a transient event involving an accreting white dwarf in a binary star system (e.g., \citealt{2008clno.book.....B}). Once the pressure and temperature at the base of the accreted envelope reach a critical level, a thermonuclear runaway is triggered on the surface of the white dwarf, leading to the ejection of at least part of the envelope. 
Typically, $10^{-7}-10^{-4}$ M$_{\odot}$ of material is ejected at velocities ranging between 500 and 5,000 km\,s$^{-1}$ \citep[e.g.,][]{Payne-Gaposchkin_1957,Gallaher_etal_1978, Yaron_etal_2005}. Remnants of the accreted envelope remain on the white dwarf's surface and continue nuclear burning for weeks to years after the thermonuclear runaway ends, bathing the ejecta with luminous ionizing radiation from within ($\sim 10^{38}$ erg s$^{-1}$; 
\citealt{Wolf_etal_2013}). Early in the nova's evolution, the ejecta are optically thick, and as the thermal emission from the white dwarf diffuses through the ejecta, the nova's spectral energy distribution peaks in the optical band. As the ejecta expand, their density drops, they become more optically-thin, and the peak of the nova's spectral energy distribution moves blueward \citep{Gallagher_Code_1974}. 
When the white dwarf is finally revealed, the nova is considered a supersoft X-ray source (emitting at photon energies $<0.5$\,keV), and it will remain in this state for days--years, until the residual fuel is all burnt \citep{Krautter_2008,2011ApJS..197...31S,Page_Osborne_2014,Osborne_2015}. After a period of time, accretion will resume and the process restarts. All novae are theorized to recur, but some novae have been observed to erupt more than once during our observational records; these are known as recurrent novae.

\par
The discovery of GeV $\gamma$-rays from nova V407~Cyg with the Large Area Telescope (LAT) on the \textit{Fermi Gamma-Ray Space Telescope} (henceforth \emph{Fermi}) has opened the door for a whole new realm of nova research \citep{2010Sci...329..817A}. At first, the $\gamma$-rays were thought to be the result of the ejecta interacting with the dense wind of V407~Cyg's Mira giant companion, and not a feature of typical nova systems \citep[e.g.,][]{2011MNRAS.410L..52M, Nelson_etal_2012}. However, the discovery of $\gamma$-rays from V959~Mon, V1324~Sco, and V339~Del with \textit{Fermi}-LAT in the following years revealed that V407~Cyg was not a singular case \citep{Ackermann+14}. Unlike V407~Cyg, these systems contain main-sequence companions, so the $\gamma$-rays could not be coming from the ejecta interacting with a dense circumbinary medium. 

\par
Since 2013, even more novae with main sequence companions have been detected in the GeV $\gamma$-ray band. These observations reveal that shocks are common in nova eruptions and that they are energetically important \citep{Li_etal_2017,Aydi2020}. As the majority of \textit{Fermi}-detected novae have dwarf (rather than giant) companions and low-density circumbinary material, the $\gamma$-ray emitting shocks must be internal to the nova ejecta. From high-resolution radio imaging of the nova V959~Mon, it was found that these shocks may occur at the interface between a slow, dense, equatorial torus and a fast biconical wind \citep{Chomiuk_etal_2014}. The shocks produced at these interfaces accelerate particles to relativistic speeds via the diffusive shock mechanism and lead to GeV $\gamma$-ray emission \citep{2015MNRAS.450.2739M}. These internal shocks have velocities $\sim$1000 km s$^{-1}$, and consequently heat the post-shock gas to temperatures of $\sim 10^7$ K, which emits relatively hard ($\gtrsim$1\,keV; compared to the supersoft component) X-rays. Even before $\gamma$-rays were detected in novae, hard X-ray emission was observed and interpreted as an indication of shock interaction (e.g., \citealt{2001ApJ...551.1024M,2008ApJ...677.1248M}).

In the last two decades, the X-ray Telescope (XRT; \citealt{2005SSRv..120..165B}) on the \emph{Neil Gehrels Swift Observatory} (hereafter \textit{Swift}; \citealt{2004ApJ...611.1005G}) has been instrumental in providing observations for novae in the 0.3\,--\,10\,keV band at relatively high cadence (e.g., \citealt{2019arXiv190802004P}).
For example, V407 Cyg showed hard X-ray emission during its first months of evolution concurrent with the $\gamma$-ray producing phase; as with the $\gamma$-rays, this is likely a result of the nova ejecta interacting with the secondary's wind \citep{2010Sci...329..817A, Orlando_Drake12, 2012ApJ...748...43N}. 

The hard X-ray behavior of classical novae with main sequence companions is less clear, especially while $\gamma$-rays are being detected.  
\textit{Swift} observed V1324~Sco while GeV $\gamma$-rays were detected, but failed to detect any X-rays \citep{Finzell_etal_2018}. 
Based on this non-detection, \citet{2014MNRAS.442..713M} theorized that the X-ray emission from classical novae during the $\gamma$-ray period would be absorbed by the initially dense ejecta.
These absorbed X-rays are then reprocessed and re-emitted as UV and optical photons, contributing to the luminosity in those bands (supporting this hypothesis, correlated $\gamma$-ray and optical light curves have been observed in two novae to date; \citealt{Li_etal_2017,Aydi2020}). Once the ejecta expand enough and the optical depth decreases, X-rays are allowed through.

Similar to V1324~Sco, there are hints from other novae that 1--10\,keV X-rays were not detectable by \emph{Swift} until $\gtrsim$1 month after eruption (e.g., \citealt{2016A&A...590A.123S}, \citealt{2018ApJ...853...27M}), but the X-ray light curves were not explicitly discussed in the context of $\gamma$-rays and shocks. 
Interestingly, harder X-rays ($>$10\,keV) from novae have begun to be detected with \emph{NuSTAR} concurrent with $\gamma$-rays
\citep{Nelson_etal19, 2020NatAs...4..776A, 2020MNRAS.497.2569S}, but at surprisingly low fluxes (the implications of these observations will be discussed in Section \ref{sec:disc_conc}). Despite the rapid response and agility of \textit{Swift} that make it ideal for observations during the early weeks of nova eruptions (when novae are bright in GeV $\gamma$-rays), no systematic study has been carried out of \emph{Swift}-XRT observations of $\gamma$-ray detected novae.
It is the goal of this paper to test if all classical novae are faint in the 1--10 keV X-ray band during $\gamma$-ray detection.



\begin{deluxetable*}{lccccc}[t]
\tabletypesize{\small}
\tablewidth{0 pt}
\tablecaption{Characteristics of $\gamma$-ray detections of  novae (2010--2018). 
\label{table:gamma}}
\tablehead{Nova & Time$_{\rm \gamma-ray\ start}$ & Time$_{\rm \gamma-ray\ end}$ & $\gamma$-ray Flux & Photon index & Reference\\
 & (MJD) & (MJD) & (10$^{-7}$ photon s$^{-1}$ cm$^{-2}$) & }
\startdata
V392 Per & 58238\tablenotemark{b} & 58246 & $2.2\pm0.4$ & $2.0\pm0.1$ & 10,11\\
V906 Car & 58216\tablenotemark{a} & 58239--58250\tablenotemark{a} & $12.2\pm0.4$ & $2.04\pm0.02$ & 9 \\
V357 Mus & 58129 & 58156 & $1.3\pm0.2$ & $2.2\pm0.1$ & 8, This work\\
V549 Vel & 58037 & 58070 & $0.4\pm0.2$ & $1.8\pm0.2$ & 6,7 \\
V5856 Sgr & 57700 & 57715 & $4.6\pm0.5$ & $2.11\pm0.05$ & 5 \\
V5855 Sgr & 57686 & 57712 & $3.0\pm0.8$ & $2.26\pm0.12$ & 4 \\
V407 Lup & 57657 & 57660 & $1.6\pm0.7$ & $2.2\pm0.3$ & 3, This work \\
V5668 Sgr & 57105 & 57158 & $1.1\pm0.2$ & $2.42\pm0.13$ & 2 \\ 
V1369 Cen & 56634 & 56672 & $2.5\pm0.4$ & $2.37\pm0.09$ & 2 \\
V339 Del & 56520 & 56547 & $2.3\pm0.3$ & $2.26\pm0.08$ & 1 \\
V959 Mon & 56097 & 56119 & $4.8\pm0.6$ & $2.34\pm0.09$ & 1 \\
V1324 Sco & 56093 & 56110 & $5.9\pm0.9$ & $2.16\pm0.09$ & 1 \\
V407 Cyg & 55265 & 55287 & $5.8\pm0.6$ & $2.11\pm0.06$ & 1 \\
\enddata
\tablenotetext{}{References: 1= \cite{Ackermann+14}; 2= \cite{2016ApJ...826..142C}; 3= \cite{2016ATel.9594....1C}; 4= \cite{Nelson_etal19}; 5= \cite{Li_etal_2017}; 6= \cite{2017ATel10977....1L}; 7= \cite{2020arXiv201010753L}; 8= \cite{2018ATel11201....1L}; 9= \cite{Aydi2020}; 10= \cite{2018ATel11590....1L}; 11= Blochwitz, Linnemann et al.\ 2020, in prep.}
\tablenotetext{a}{Due to \emph{Fermi}-LAT downtime, the start time of $\gamma$-ray detection for V906~Car was not captured, and the end time is only constrained to be within a date range. The $\gamma$-ray flux is calculated over MJD 58216--58239.}
\tablenotetext{b}{Due to \emph{Fermi}-LAT downtime, data are not available for MJD 58224-58238. When observations resumed on MJD 58238, V392 Per was immediately detected. The $\gamma$-ray flux is calculated over MJD 58238--58246.}
\end{deluxetable*}

\par Previous studies have been carried out on large collections of novae in the supersoft X-ray phase using observations from \emph{Swift}-XRT,
including \citet{2007ApJ...663..505N,2011ApJS..197...31S}, and \citet{2019arXiv190802004P}. However, systematic studies of the harder X-ray component, or the X-ray behavior of $\gamma$-ray detected novae, are lacking. 
In this paper, we present a systematic study of 13 Galactic novae which have been detected by \textit{Fermi}-LAT between 2010 and 2018 and have been observed by \textit{Swift}-XRT. In Section~\ref{sec:style} we discuss the sample selection and the multi-wavelength properties of the novae in our sample. \textit{Fermi}-LAT data for most novae in our sample have already been published, but we present the first $\gamma$-ray analysis of novae V407~Lup and V357~Mus.
In Section~\ref{subsec:x-ray light ces} we present the \textit{Swift}-XRT observations, emphasizing the hard X-ray emission during the $\gamma$-ray detection phase. 
In Section~\ref{sec:disc_conc} we discuss what can be learned about nova shocks from observations concurrent with $\gamma$-rays, and in Section~\ref{sec:conc} we conclude.

\section{Our Sample of $\gamma$-ray Detected Novae} \label{sec:style}

\subsection{Sample Selection}
In this paper, we analyze all Galactic novae observed by \emph{Fermi}-LAT between 2010 and 2018 that have a time-integrated detection $\geq3 \sigma$ significance over the period of $\gamma$-ray emission. Details of the sample are listed in Table \ref{table:gamma}. Despite hints that they produced $\gamma$-ray emission, we do not include novae V745~Sco, V697~Car, or V1535~Sco in our sample because their \emph{Fermi}-LAT detections were $<$3$\sigma$ significance \citep{Franckowiak_etal_2018}.

\subsection{$\gamma$-ray Properties}\label{sec:gamma}
Parameterizations of the $\gamma$-ray light curves for our 13 novae are provided in Table \ref{table:gamma}, taken from references listed therein. Time$_{\rm \gamma-ray\ start}$ and time$_{\rm \gamma-ray\ end}$ denote the time range during which $\gamma$-rays are detected at $>$2$\sigma$ significance when binning \emph{Fermi}-LAT light curves with 1-day cadence. The $\gamma$-ray flux column lists the average flux over this time period, fitting a single power law to the data over the energy range $>$100 MeV. Table \ref{table:gamma} also lists the photon index for a single power law fit to the \emph{Fermi}-LAT data with energy $>$100 MeV: 
\begin{equation}
    \dfrac{dN}{dE} \propto E^{-\Gamma},
\end{equation}
where $N$ is the number of photons, $E$ is the photon energy, and $\Gamma$ is the photon index. Although a single power law may not be the most physically motivated model, it is the simplest (most justified in cases of low S/N), and most widely quoted in studies of the various novae. It is sufficient for estimating $\gamma$-ray luminosities to the precision required for this study; modelling with a more complex exponentially-cutoff power law spectrum yields fluxes 75--85\% that of a simple power law \citep{Ackermann+14}.

The $\gamma$-ray detections of novae V407~Lup and V357~Mus have been announced in \cite{2016ATel.9594....1C} and \cite{2018ATel11201....1L} respectively, but a full analysis of their light curves has not yet been published. We therefore provide this analysis here, in the following sub-sections.

\begin{figure*}[t]
    \centering
    \includegraphics[width=0.8\textwidth]{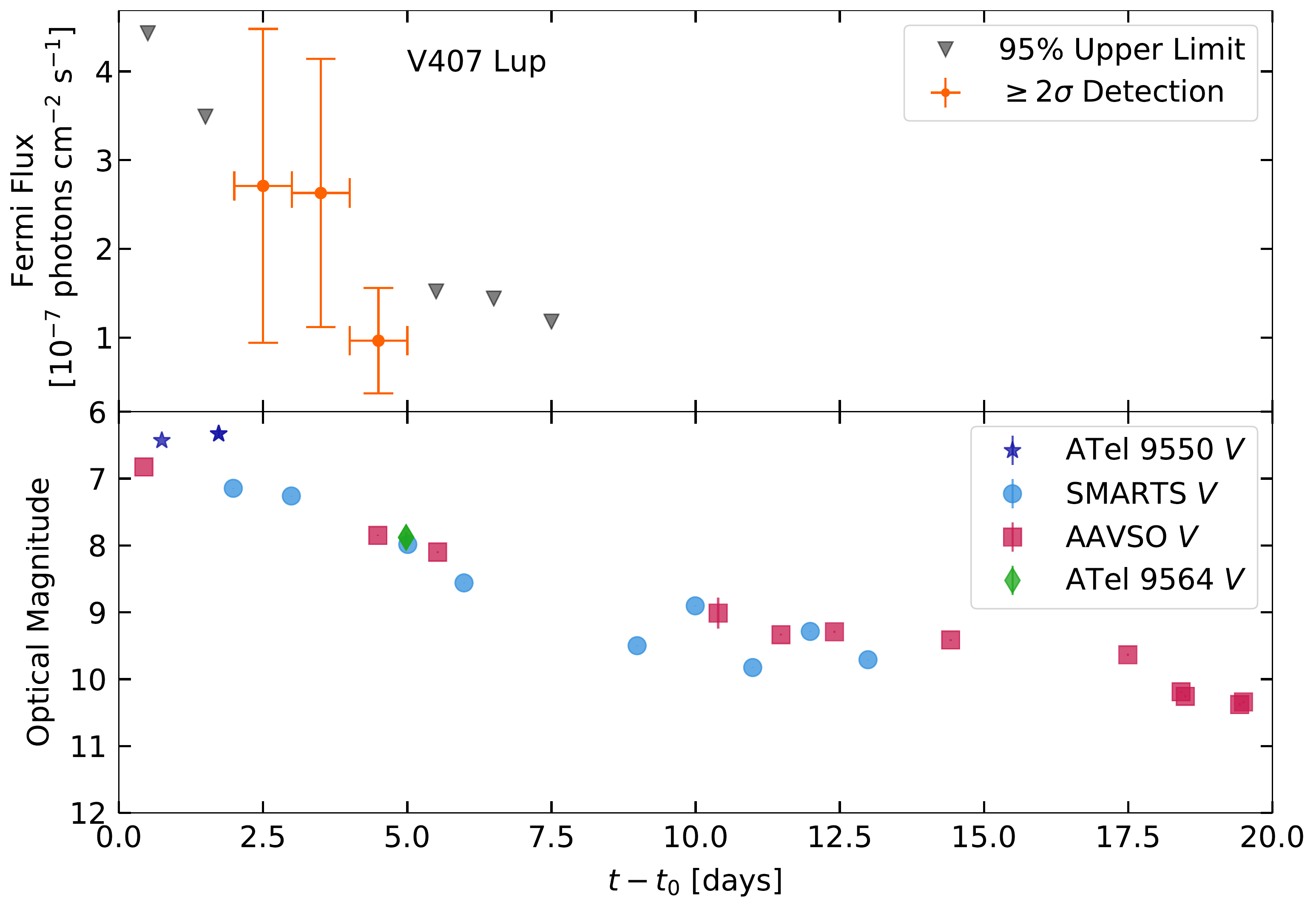}
    \caption{Top panel: the $\gamma$-ray light curve ($>$100 MeV) of V407 Lup. Bottom panel: the $V$-band optical light curve of V407 Lup. $t_0$ is taken to be the time of discovery, 2016 Sep 24.}
    \label{fig:V407 Lup gamma light curves}
\end{figure*}

\subsubsection{\textit{Fermi}-LAT data reduction}

We downloaded the LAT data (Pass 8, Release 3, Version 2 with the instrument response functions of \texttt{P8R3\_SOURCE\_V2}) from the data server at the \textit{Fermi Science Support Center} (FSSC). 
For data reduction and analysis, we used \texttt{fermitools} (version 1.0.5) with \texttt{fermitools-data} (version 0.17)\footnote{\url{https://fermi.gsfc.nasa.gov/ssc/data/analysis/software/}}. For data selection, we used a region of interest $14^\circ$ on each side, centered on the nova.
Events with the class \texttt{evclass=128} (i.e., SOURCE class) and the type \texttt{evtype=3} (i.e., reconstructed tracks FRONT and BACK) were selected. We excluded events with zenith angles larger than $90^\circ$ to avoid contamination from the Earth's limb. The selected events also had to be taken during good time intervals, which fulfils the \texttt{gtmktime} filter \texttt{(DATA\_QUAL$>$0)\&\&(LAT\_CONFIG==1)}. 

Next, we performed binned likelihood analysis on the selected LAT data. For each nova, a $\gamma$-ray emission model for the whole region of interest was built using all of the 4FGL cataloged sources located within $20^\circ$ of the optical position \citep{2019arXiv190210045T}. 
As the two novae were the brightest $\gamma$-ray sources in the fields (within at least 5 degrees according to the preliminary results), we only freed the normalization parameters for those cataloged sources located less than 1 degree from the targets.
In addition, the Galactic diffuse emission and the extragalactic isotropic diffuse emission were included by using the Pass 8 background models \texttt{gll\_iem\_v07.fits} and \texttt{iso\_P8R3\_SOURCE\_V2\_v1.txt}, respectively, which were allowed to vary during the fitting process.


\begin{figure*}[t]
    \centering
    \includegraphics[width=0.8\textwidth]{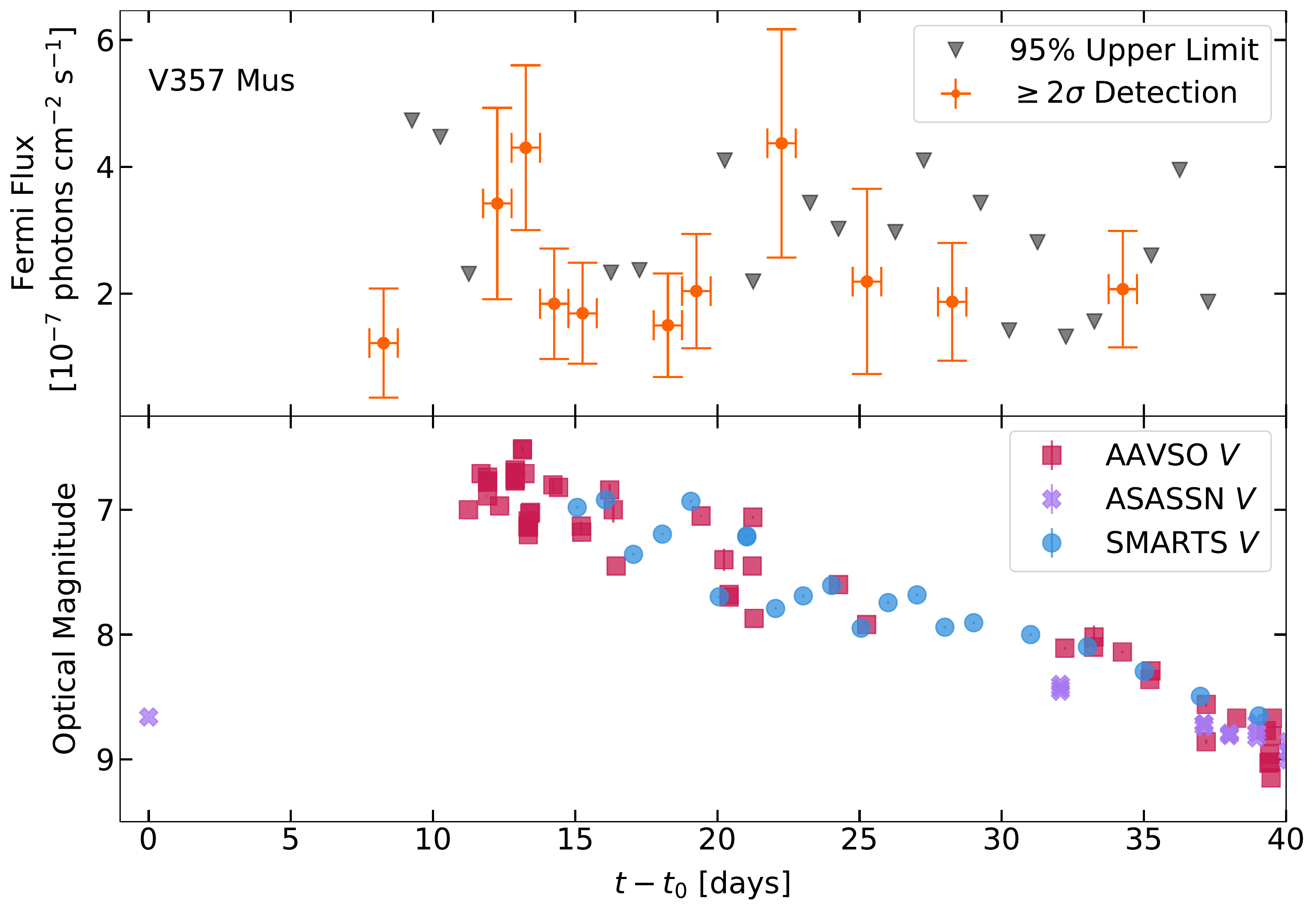}
    \caption{Top panel: the $\gamma$-ray light curve ($>$100 MeV) of V357 Mus. Bottom panel: the $V$-band optical light curve of V357 Mus. $t_0$ is taken to be the time of first observation in outburst, 2018 Jan 3.}
    \label{fig:V357 Mus gamma light curve}
\end{figure*}

\subsubsection{$\gamma$-rays from V407~Lup} \label{sec:lup}

Nova V407 Lup (ASASSN-16kt) was discovered by the All-Sky Automated Survey for Supernovae (ASAS-SN) on 2016 September 24.0 UT at $V$ = 9.1 (\citealt{2016ATel.9538....1S, 2018MNRAS.480..572A}). The nova was first detected in $\gamma$-rays on the same day as its optical discovery but did not reach a significance of $\geq$3$\sigma$ until the next day, when the detection significance reached $4\sigma$ (with a Test Statistic $TS = 16.6$). 
The average flux over the detection duration was $(1.6 \pm 0.7)\,\times\,10^{-7}$ photon s$^{-1}$\,cm$^{2}$. A single power-law fit gives a photon index of $2.2\pm0.3$. 

The $\gamma$-ray light curve, shown in the top panel of Figure~\ref{fig:V407 Lup gamma light curves}, shows a decrease in flux, completely fading below the LAT detection limit by 5 days after discovery. This makes V407~Lup's $\gamma$-ray duration the shortest known to date for a nova (Table \ref{table:gamma}). The bottom panel of Figure \ref{fig:V407 Lup gamma light curves} shows the optical behavior of V407 Lup in the V-band during and shortly after the $\gamma$-ray period.
This light curve is constructed from publicly available photometry from the American Association of Variable Star Observers (AAVSO; \citealt{Kafka20}) and the Stony Brook/SMARTS Atlas (SMARTS photometry can be found at Ref.\footnote{\url{http://www.astro.sunysb.edu/fwalter/SMARTS/NovaAtlas/}}; \citealt{Walter_etal_2012}), along with  \citet{2016ATel.9550....1C} and \citet{2016ATel.9564....1P}.
The optical light curve rapidly declines alongside the $\gamma$-rays, exhibiting the shortest $t_2$ value of any nova in our sample (Table \ref{table:characteristics}).
The light curve shows what could possibly be a lag in the $\gamma$-ray emission compared to the optical. However, the date of the optical peak as estimated by \citet{Aydi_etal_2018} is MJD 58656.4, implying that the first 3$\sigma$ $\gamma$-ray detection lags the optical peak by only 0.6 days. Based on the cadence of the optical and $\gamma$-ray observations, which is around 0.5 days, this delay may be insignificant. A delay between the optical and $\gamma$-ray emission, if it exists, would have significant implications on our understanding of shock formation and $\gamma$-ray emission in novae (see e.g., \citealt{2015MNRAS.450.2739M} and \citealt{Aydi2020}).

\subsubsection{$\gamma$-rays from V357~Mus} \label{sec:mus}

Nova V357 Mus was discovered in the optical on 2018 Jan 14.5 UT at $\sim$7 mag \citep{cbet4473}. It was first detected in $\gamma$-rays eight days later \citep{2018ATel11201....1L}. The average $>$100 MeV flux over the detection period was $(1.3\pm0.2) \times 10^{-7}$ photon s$^{-1}$ cm$^{2}$, and the photon index from fitting a single power law was $\Gamma = 2.2\pm0.1$. The detection significance was $10\sigma$ over this period (with a Test Statistic $TS = 98.6$).

The $\gamma$-ray light curve is shown in the top panel of Figure~\ref{fig:V357 Mus gamma light curve}. There may be variability of a factor of $\sim$2 in the light curve, but the low S/N makes it challenging to confidently measure this variability. The corresponding optical light curve is shown in the bottom panel of Figure \ref{fig:V357 Mus gamma light curve}, with data from ASAS-SN  \citep{Shappee_etal_2014}, the Stony Brook/SMARTS Atlas, and AAVSO. The nova was detected by ASAS-SN on the rise to optical maximum but quickly became so bright that it saturated the detectors. Observations resumed around day 10 when amateur observers found the nova and began taking data \citep{cbet4473}. The nova likely reached a magnitude brighter than 6\,mag at optical maximum  (which was sometime between 0--10 days after the ASAS-SN pre-maximum detection). While the cadence and S/N of the $\gamma$-ray and optical light curves are not high enough to confirm, this nova may show evidence of correlated variation between the optical and $\gamma$-ray light curves, similar to the two brightest $\gamma$-ray novae V906 Car \citep{Aydi2020} and V5856~Sgr \citep{Li_etal_2017}. 

\begin{longrotatetable}
\begin{deluxetable*}{lcccccccccc}
\tabletypesize{\small}
\tablewidth{0 pt}
\tablecaption{\label{table:characteristics} Nova Properties}
\tablehead{Name & $t_0$\tablenotemark{a} & $t_0$\tablenotemark{a} & Discovery Mag\tablenotemark{b} & $V_{max}$ & Dust? & $t_2$ & Spec.\ Class & FWHM & Distance & N(H) \\
& (MJD) & (Date, UT) &  (mag) & (mag) & (Y/N) & (days) &  & (km s$^{-1}$) & (kpc) & ($10^{21}$ cm$^{-2}$)}
\startdata
V392~Per & 58237.47 (1) & 2018-04-29.47 (1) & $\sim$6.2 (1) & 5.6 (2) & - & 3 (29) & Fe II (2) & 4700$\pm200$ (2) & 4.1$^{+2.3}_{-0.4}$ (25) &3.4$\pm$0.4 (2)\\
V906~Car & 58193.03 (28) & 2018-03-16.03 (28) & $<$10\tablenotemark{c} (3) & $\sim$5.9 (28) & Y (28) & 44$\pm2$ (28) & Fe II (28) & 1500$\pm100$ (2) & 4.0$\pm1.5$ (28) & 3.1$\pm$0.4 (2) \\
V357~Mus & 58121.24 (4) & 2018-01-3.24 (4) &  7.0 (5) & 7.0 (5) & - & 40$\pm5$ (2) & Fe II (5) & 1200$\pm100$ (2) & 3.2$\pm0.5$ (2) & 4.2$\pm$0.8 (2)\\
V549~Vel & 58020.39 (6) & 2017-09-24.39 (6) & $\sim$11.3 (6)  & 9.1 (2) & - & 90 (2)  & Fe II (7) & 2300$\pm$200 (2) & $>$4 (2) & 9.0$\pm$1.0 (2)\\
V5856~Sgr & 57686.02 (8) & 2016-10-25.02 (8) & $\sim$13.7 (9) & 5.4 (8) & - & 10 (8) & Fe II (8) & 1600$\pm100$ (2) & 2.5$\pm0.5$ (2) & 3.1$\pm$0.4 (2)\\
V5855~Sgr & 57681.38 (11) & 2016-10-20.84 (11) & 10.7\tablenotemark{d} (11) & 7.5 (11) & - & 17$\pm2$ (2) & Fe II (12) & 200$\pm$200 (2) & 4.5 (11) & ---\\
V407~Lup & 57655.00 (13) & 2016-09-24.00 (13) &  $\sim$9.1 (13) & $<$5.6 (14) & N (2) & 3$\pm$1 (5) & He/N (14) & 2900$\pm$100 (2) & 4.2$\pm$0.5 (2) & 9.0$\pm$1.2 (2)\\
V5668~Sgr & 57096.63 (15) & 2015-03-15.63 (15) & 6.0\tablenotemark{d} (15) & 4.4 (16) & Y (17) & 75$\pm$2 (2) & Fe II (2) & 1300$\pm$100 (2) & 2.8$\pm$0.5 (2) & 5.9$\pm$0.8 (2)\\
V1369~Cen & 56628.69 (18) & 2013-12-2.69 (18) & 5.5\tablenotemark{d} (18) & $\sim$3.3 (2) & Y (2) & 40$\pm$2 (2) & Fe II (2) & 1200$\pm$100 (2) & 1.0$\pm$0.4 (2) & 0.6$\pm$0.1 (2)\\
V339~Del & 56518.58 (19) & 2013-08-14.58 (19) & 6.8\tablenotemark{d} (19) & $\sim$4.3 (2) & N (2) & 11$\pm$1 (2) & Fe II (2) & 1700$\pm$100 (2) & 4.9$\pm$1 (2) & 1.7$\pm$0.4 (2)\\
V959~Mon & 56097.00 (20) & 2012-06-19.00 (20) & $\sim$9.9 (2) & N/A\tablenotemark{f} (27) & N (2) & 10 (2) & He/N (21) & 2000$\pm$200 (2) & 1.4$\pm$0.4 (26) & 3.4$\pm$0.4 (2)\\
V1324~Sco & 56069.80 (22) & 2012-05-22.80 (22) & 18.5\tablenotemark{e} (23) & 9.8 (2) & Y (22) & 24 (29) & Fe II (22) & 1900$\pm$200 (2) & $>$6.5 (29) & 10.1$\pm$0.7 (2)\\
V407~Cyg & 55265.81 (24) & 2010-03-10.81 (24) & 6.8\tablenotemark{d} (24) & 7.1 (24) & - & 5.9 (24) & He/N (24) & 1400$\pm$100 (2) & 3.4$\pm$0.5 (2) & 5.6$\pm$0.8 (2)\\
\enddata
\tablenotetext{a}{Date of first observation in eruption.}
\tablenotetext{b}{$V$ band, unless otherwise noted.}
\tablenotetext{c}{Image was saturated.}
\tablenotetext{d}{Image was obtained in an unfiltered optical band.}
\tablenotetext{e}{Image was obtained in the $I$ band.}
\tablenotetext{f}{Optical maximum was during solar conjunction, so was missed}
\tablenotetext{}{References (1) \citet{Munari_etal_2018}; (2) This work; (3) \citet{Stanek_etal_2018}; (4) ASAS-SN data 
\citet{Walter_etal_2018}; (5) \citet{Aydi_etal_2018}; (6) \citet{Stanek_etal_2017}; (7) \citet{Luckas_etal_2017}; (8) \citet{Li_etal_2017}; (9) AAVSO Alert 561; (10) \citet{Munari_etal_2017}; (11) \citet{Nelson_etal19}; (12) \citet{Luckas_etal_2016}; (13) \citet{2016ATel.9538....1S}; (14) \citet{2018MNRAS.480..572A}; (15) \citet{2016ApJ...826..142C}; (16) \citet{2018ApJ...858...78G}; (17) \citet{2015ATel.7748....1B}; (18) \citet{2013CBET.3732....3W}; (19) \citet{2013AAN...489....1W}; (20) \citet{Ackermann+14}; (21) \citet{2013ATel.4709....1M}; (22) \citet{Finzell_etal_2018}; (23) \citet{2012ATel.4157....1W}; (24) \citet{2011MNRAS.410L..52M};  (25)
\citet{2018MNRAS.481.3033S};  (26)
\citet{2015ApJ...805..136L}; (27) \citet{2014ASPC..490..217S}; (28) \citet{Aydi2020}; (29) \citet{Chochol_etal_2020}; (30) \citet{Finzell_etal_2015}}
\end{deluxetable*}
\end{longrotatetable}

\subsection{Optical Properties}\label{sec:optical}
Table~\ref{table:characteristics} presents the main characteristics of the novae in our sample,
some of which are compiled from the literature (with references given in parentheses following each table entry) and others estimated for the first time here. 
It includes date of first detection in eruption ($t_0$) in MJD and UT, optical magnitude at $t_0$, peak magnitude in the $V$-band ($V_{max}$), and the time for the optical light curve to decline by two magnitudes from maximum ($t_2$). The peak magnitude and $t_2$ are determined from reports in the literature or derived in this work using publicly available photometry from the AAVSO, ASAS-SN,   and the Stony Brook/SMARTS Atlas. $t_2$ is measured as the duration between the first peak and the last time the nova reaches two magnitudes fainter than the peak.

We also list whether or not the nova formed dust based on reports in the literature or examining publicly available optical and near-infrared (NIR) photometry, particularly from SMARTS and AAVSO, to search for dust dips in the optical light and/or IR excess. For some novae, we cannot tell if the nova has formed dust or not due to lack of multi-band photometric follow-up. 

We give the spectroscopic class (\eal{Fe}{II} or He/N; \citealt{Williams92}) and the Full Width at Half Maximum (FWHM) of Balmer emission lines after optical peak. The spectroscopic classes are based on previous reports in the literature or determined based on spectra obtained around optical peak ($\lesssim t_2$). These spectra are either publicly available spectra from the Astronomical Ring for Access to Spectroscopy (ARAS\footnote{\url{http://www.astrosurf.com/aras/Aras_DataBase/Novae.htm}}; \citealt{Teyssier_2019}) or from our private database. The FWHM are measured from the same spectra by fitting a single Gaussian profile to the Balmer emission lines. Nova V959 Mon is an exception since this nova was discovered in optical 56 days after its $\gamma$-ray detection by \textit{Fermi}-LAT due to solar conjunction. The optical spectrum we use to determine the FWHM has been obtained 3 days after its optical discovery (around 60 days after optical peak, given that for most novae the $\gamma$-ray detection occurs near optical peak). 

We also use high-resolution optical spectroscopy to estimate the Galactic column density towards each nova. Again, these spectra are either from ARAS or from our private database, and are obtained near the light curve peak. We measure the equivalent widths of some diffuse interstellar bands (DIBs) and use the empirical relations of \citet{Friedman_etal_2011} to derive an estimate of $E(B-V)$. $A_V$ is then derived assuming an extinction law of $R_V$ = 3.1. This $A_V$ is converted into an absorbing column density, $N(H)$, using the relation from \citet{Bahramian_etal15}: $N(H)$ = ($2.81\,\pm\,0.13$) $\times 10^{21}\, A_V$. 

We also list distance estimates to the novae in our sample. For novae without an accurate distance estimate in the literature, we estimate the distance using our derived extinction values, along with the 3D Galactic reddening maps of \citet{Chen_etal_2019}. 

In our nova sample, one system is known to have a Mira giant secondary, namely V407~Cyg. The other 12 novae in the sample are likely systems with dwarf secondaries and will be designated as ``classical novae'' in the rest of the paper. However, it should be noted that V392~Per was recently found to have a mildly-evolved secondary star, with a binary orbital period of 3.4 days \citep{Munari+20}, implying that V392~Per may be a ``bridge'' object between embedded novae with dense circumstellar material and classical novae with low-density surroundings.

\section{X-Ray Light Curves} \label{subsec:x-ray light ces}

\subsection{\emph{Swift}-XRT observations}
The \textit{Swift}-XRT data products generator \citep{2007A&A...469..379E,2009MNRAS.397.1177E} was used to produce X-ray (0.3--10\,keV) light curves for all the novae in our sample. The same tool was also used to divide the XRT flux into soft (0.3--1.0\,keV) and hard (1--10\,keV) X-ray bands. 
Once the data products were generated, the X-ray count rates were filtered to separate the significant detections from the upper limits. Observations with less than 3$\sigma$ confidence on their count rates were considered  upper limits. We quote 3$\sigma$ upper limits throughout this paper, calculated using the uncertainty on the count rate.

In Figures~\ref{fig:V392_Per}--\ref{fig:V407_Cyg}, we present the \textit{Swift} X-ray (0.3--10\,keV) light curves of all the novae in our sample. In each figure's top panel, we plot the total XRT count rate, while the bottom panel distinguishes the light curves in the soft and hard bands. The time range of the \emph{Fermi}-LAT $\gamma$-ray detection is marked as a yellow bar, and the light curves focus on the first year following nova discovery.

There are \emph{Swift} observations concurrent with \emph{Fermi}-LAT $\gamma$-ray detections for 9 of the 13 novae in our sample. Unfortunately, for some of the \emph{Fermi}-detected novae, \emph{Swift} observations were not obtained until long after discovery, and were therefore only detected after the end of the \emph{Fermi} detection. Early observations were limited by solar conjunction for V392~Per, V959~Mon, and V549~Vel.

\begin{figure*}[t]
    \centering
    \includegraphics[width=\textwidth, height=0.41\textheight]{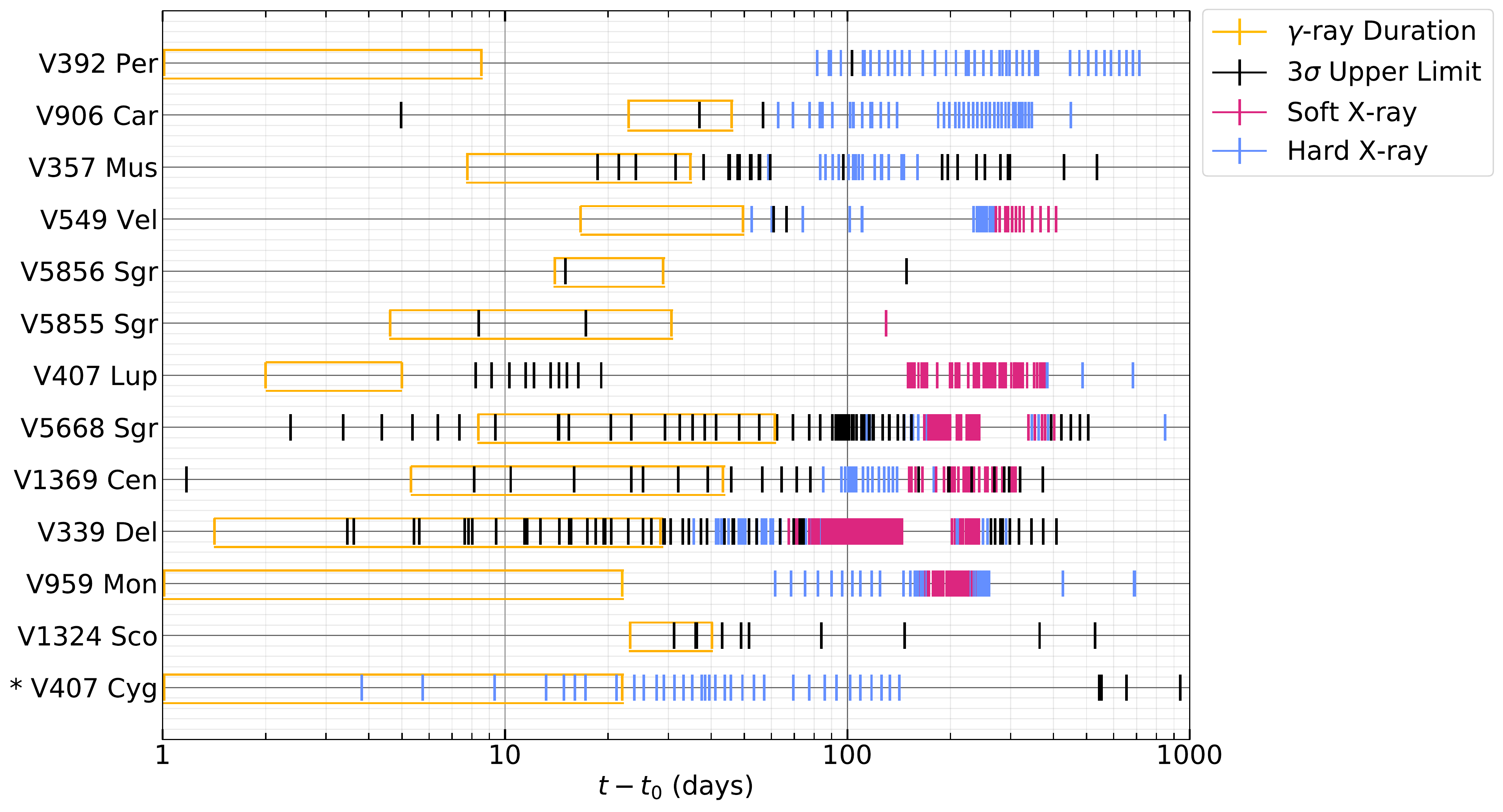}
    \caption{The evolution of the X-ray emission as a function of time since discovery for our sample of \textit{Fermi}-LAT-detected novae. Vertical tick marks represent times of \emph{Swift}-XRT observations, and are color-coded according to the type of X-ray emission that was detected. Blue ticks denote hard X-ray emission, magenta ticks are soft X-ray emission, and black ticks represent X-ray non-detections. The durations of $\gamma$-ray detections with \emph{Fermi}-LAT are denoted with yellow rectangles. V407 Cyg is marked with an asterisk to note that this system has a red giant secondary, unlike the other novae in our sample.}
    \label{fig:gamma schwarz}
\end{figure*}

Many $\gamma$-ray detected novae are very optically bright, and lead to optical loading of the XRT if observed in photon counting mode\footnote{\href{https://www.swift.ac.uk/analysis/xrt/optical_loading.php}{https://www.swift.ac.uk/analysis/xrt/optical\_loading.php}}. Therefore, some \emph{Swift}/XRT observations early in our targets' eruptions were obtained in the less sensitive windowed timing mode. This affects observations of V906~Car, V357~Mus, V1369~Cen, V5668 Sgr, and  V5856~Sgr. The supplementary online tables list information on each observation used, including the corresponding observation mode.

\subsection{Hardness ratio evolution} \label{subsec:hardness ratio evolution}
We derive the hardness ratio ($HR$) for each \textit{Swift}-XRT detection using the definition from \citet{2011ApJS..197...31S}:
\begin{equation}\label{eq:hr}
HR = (H-S)/(H+S)    
\end{equation} 
where $S$ is the count rate in the 0.3\,--\,1.0\,keV range and $H$ is the count rate in the 1\,--\,10\,keV range. We also use similar criteria as \citet{2011ApJS..197...31S} to classify the X-ray emission: we consider the X-ray emission ``hard" if HR $> -0.3$, and ``soft" if HR $< -0.3$. 

In Figure~\ref{fig:gamma schwarz} we present the evolution of the hardness ratio as a function of time since discovery for all the novae in our sample. The plot also shows the duration of the \emph{Fermi}-LAT $\gamma$-ray detection, represented as a yellow box, to compare with the \emph{Swift} X-ray observations. Non-detections, denoted as black tick marks in Figure~\ref{fig:gamma schwarz}, are defined as times when both the hard and soft bands were upper limits. If only one of the bands was detected, this epoch is counted as a detection, and the tick's color corresponds to the detected band.

The hardness ratio evolution of the novae is quite varied, but the main commonality is the lack of significant X-ray detection during the $\gamma$-ray emission period (with the exception of V407 Cyg). 

\section{Discussion}
\label{sec:disc_conc}

\subsection{The drivers of hard X-rays in novae}
\label{subsec:hard_lum}

\begin{figure*}
    \centering
    \includegraphics[width=\textwidth]{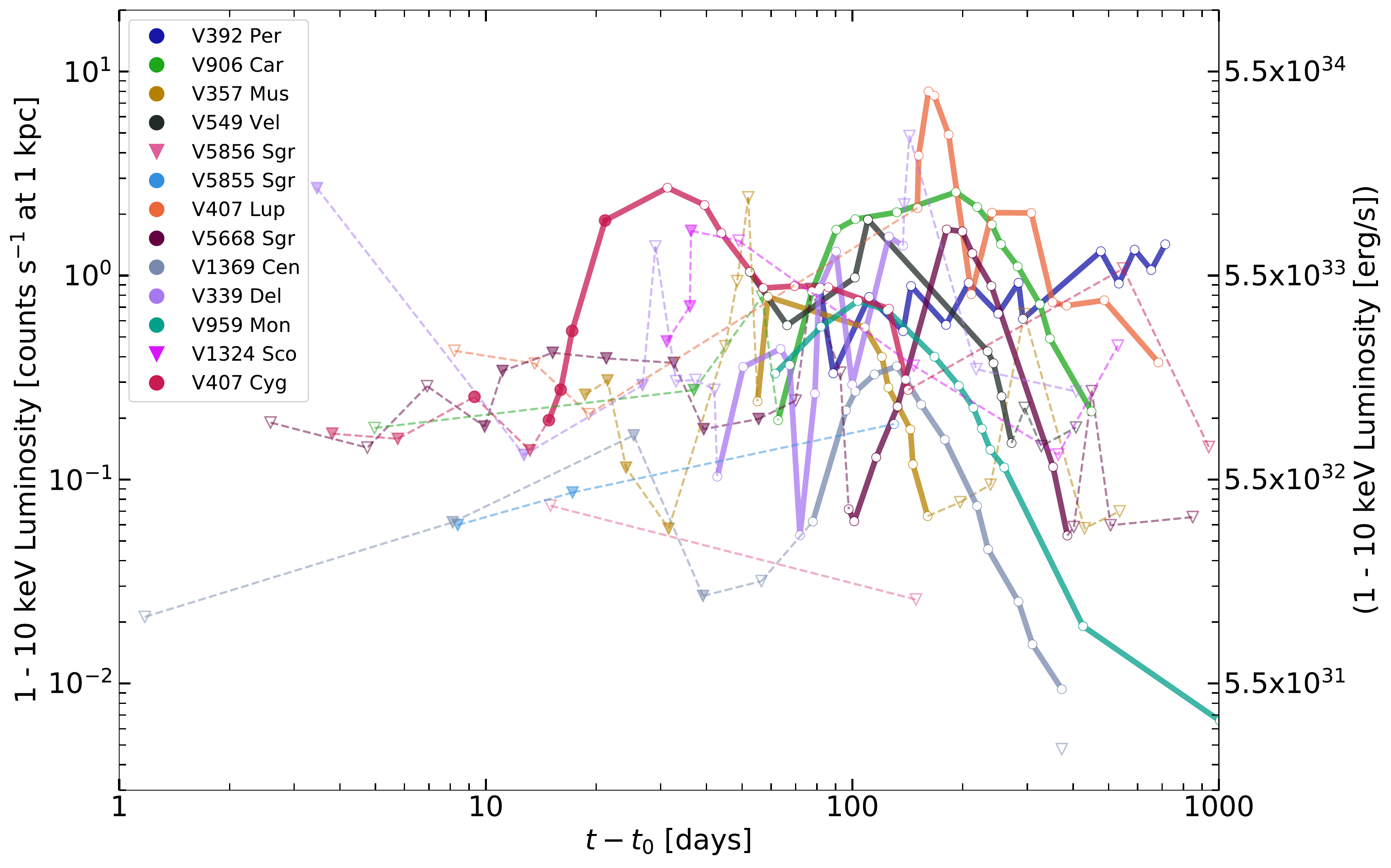}
    \caption{The luminosity in the hard X-ray band for all novae since the time the system was observed to be in eruption. When deriving these luminosities, we do not account for intrinsic absorption (e.g., absorption from the nova ejecta). Circles represent $\geq$3$\sigma$ hard X-ray detections and triangles represent upper limits.  Filled-in symbols denote concurrent hard X-ray and $\gamma$-ray emission, while open symbols were observed while $\gamma$-rays are not detected. Note that the light curves have been edited to better highlight trends.}
    \label{fig:hard_luminosity}
\end{figure*}

\par Hard X-ray emission from optically-thin plasma with a temperature of several keV has long been observed in novae
and is commonly attributed to shocks \citep[e.g.,][]{1994MNRAS.271..155O}. Figure \ref{fig:hard_luminosity} compares the luminosity light curves in the 1--10\,keV range for our nova sample. They have been smoothed to highlight bulk features; see Figures~\ref{fig:V392_Per}--\ref{fig:V407_Cyg} for full-cadence light curves. Distances are assumed as listed in Table~\ref{table:characteristics}. Count rates are corrected for foreground absorbing columns consistent with the intervening interstellar medium, using the $N(H)$ values quoted in Table~\ref{table:characteristics}. We do not account for intrinsic absorption (e.g., absorption from the nova ejecta). To make a rough conversion of the \emph{Swift}-XRT count rate to X-ray flux, we assume a 5 keV thermal bremsstrahlung model (yielding the scale in units of erg s$^{-1}$ on the right y axis). The unabsorbed fluxes were then corrected for absorption using WebPIMMs as described above and then converted to luminosities scaled at a distance of 1 kpc.

\par This figure updates a similar plot from \citet{2008ApJ...677.1248M}, with the goal of exploring the luminosities and durations of the hard X-ray emission from shocks. Figure 1 in \citet{2008ApJ...677.1248M} shows hard X-ray luminosity as a function of time for 16 novae, but most had very limited time coverage so the duration of hard X-rays was unclear. In our sample, we see that typically the hard X-rays become detectable 1--2 months after the start of eruption, and last several months to $\sim$a year. The notable exception is V407 Cyg, whose hard X-rays evolve much faster, starting shortly after day 10 (concurrent with $\gamma$-ray detection).
This rapid evolution may be attributable to interaction with circumbinary material around the secondary (\S \ref{sec:v407}); it is notable that other novae with giant companions, RS~Oph \citep{Sokoloski+06, Bode+06, 2008ApJ...677.1248M}, V745~Sco \citep{2015MNRAS.454.3108P}, and V3890~Sgr \citet{2020arXiv201001001P} were all detected in hard X-rays from the first pointed observations. However, RS Oph's last eruption occurred before the launch of Fermi, so we do not have information on its $\gamma$-ray evolution. 

\par The 1--10 keV X-ray luminosities of novae in Figure \ref{fig:hard_luminosity} peak at $10^{33}-10^{34}$ erg s$^{-1}$. While we expect the bulk of the 1--10 keV luminosity to originate from shocked optically thin gas, in some cases it may be dominated by the hard tail of the supersoft component. For example, for moderate absorbing columns $N(H) \lesssim 10^{22}$ cm$^{-2}$, as expected for the Galactic foreground (Table \ref{table:characteristics}), a $10^{37.5}$ erg s$^{-1}$ blackbody of temperature $T_{BB} =$ 90 eV produces $\sim$30 times as many counts in the 1--10 keV band\footnote{\url{https://heasarc.gsfc.nasa.gov/cgi-bin/Tools/w3pimms/w3pimms.pl}} as a $10^{34}$ erg s$^{-1}$  bremsstrahlung component of temperature 5 keV. Such a hot supersoft component is only expected for a near-Chandrasekhar mass white dwarf (e.g., \citealt{Osborne_etal_2011}), and contamination of the 1--10 keV band depends sensitively on the temperature of the supersoft source. A more typical white dwarf ($T_{BB} \approx$ 60 eV; \citealt{Wolf_etal_2013}) of a similar luminosity contaminates the 1--10 keV band orders of magnitude less severely, contributing $\lesssim$30\% of the 1--10 keV flux.

\par A detailed analysis of when the supersoft source contributes significantly to the 1--10 keV band would require spectral fitting of the \emph{Swift}-XRT data and is outside the scope of this paper. However, in the case of a moderate absorbing column, the hardness ratio is a powerful discriminant. For $N(H) = 5 \times 10^{21}$ cm$^{-2}$, the count rate from a supersoft source should be $\gtrsim 10 \times$ higher in the 0.3--1 keV band compared to  the 1--10 keV band. For example, the 1--10 keV X-rays observed from V407~Lup starting around day 150 are likely the hard tail of the supersoft source because the concurrent 0.3--1 keV X-rays are so much brighter (Figure \ref{fig:V407_Lup}). At higher absorbing columns, the ratio of soft-to-hard X-rays from a supersoft source will be lower. In the case of V339 Del and V1369 Cen, some of the early 1--10 keV X-rays (day $\sim$50 and day $\sim$80, respectively) could be attributed to the supersoft source beginning to emerge from the absorbing nova ejecta, as the 0.3--1 keV flux is increasing during this time (Figures \ref{fig:V1369_Cen} and \ref{fig:V339_Del}). However, in many of the novae studied here, the 1--10 keV flux is significantly brighter than the 0.3--1 keV flux and relatively stable in time (e.g., Figures \ref{fig:V906_Car}, \ref{fig:V357_Mus}, \ref{fig:V959_Mon}), implying that the early hard X-rays really are emitted from hot shocked gas.

\par It is also possible that accretion could be a source of hard X-rays, particularly at late times. This is mainly true for systems with highly magnetized white dwarfs ($B>10^6$\,G), such as intermediate polars. In such systems, accretion is channeled by the strong magnetic field lines into an accretion column which then slams onto the white dwarf surface at high speeds, increasing the surface temperature and leading to hard X-ray emission (see~\citealt{Warner_1995} for a review). Per example, the hard X-ray emission in Nova V407~Lup around 350 days after eruption is probably due to accretion resuming on the surface of the white dwarf (Figure \ref{fig:V407_Lup}). This nova occurred in an intermediate-polar system where the white dwarf is highly magnetized (see \citealt{2018MNRAS.480..572A} for more details). The luminosity of hard X-ray emission in intermediate polars is usually $\sim$ 10$^{31}$\,--\,10$^{34}$\,erg\,s$^{-1}$ \citep{Patterson_etal_1994,Pretorius_Mukai_2014}, which is consistent with the X-ray luminosity of V407~Lup around 2 years after eruption.   

Nova V392~Per also shows hard X-ray emission, which is peculiarly constant over a period of more than 250 days (see also Murphy-Glaysher et al. in prep for a more detailed examination of the \textit{Swift} data). As previously mentioned, this nova was recently found to have a mildly-evolved secondary star \citep{Munari+20}. However, the origin of this constant and extended hard X-ray emission is not clear. While it could be accretion related, it is less likely to be due to shock interaction within the ejecta at this late stage. After the 1998 eruption of nova 
V2487~Oph, which is characterized by a $\sim$ 1 day orbital period \citep{Anupama_2013}, \citet{Hernanz_Sala_2002} found hard X-ray emission more than two years after the eruption with comparable luminosity to that of nova V392~Per ($\sim10^{33}$\,erg\,s$^{-1}$). \citet{Hernanz_Sala_2002} attributed this 
late X-ray emission to accretion resuming on the white dwarf. In addition, \citet{Orio_etal_2001}'s study of \textit{ROSAT} observations of a large number of novae identified late X-ray emission from several novae during quiescence, which they attributed to accretion. 


\subsection{Novae with dwarf companions are not detected in 1--10\,keV X-rays concurrent with $\gamma$-rays}
\label{disc_1}

\par The internal shocks responsible for accelerating particles to relativistic speeds and producing $\gamma$-ray emission have velocities of $\sim$few thousand km\,s$^{-1}$ and are expected to heat the post-shock gas to X-ray temperatures ($\sim 10^7$\,K; \citealt{2015MNRAS.450.2739M}). Therefore, it is surprising that we do not detect \emph{Swift} X-ray emission concurrent with GeV $\gamma$-rays among the classical novae in our sample. Nine of the novae presented here have \textit{Swift}-XRT observations during their \emph{Fermi}-LAT detections, and all except V407~Cyg show no X-ray emission during this period. The other four novae did not have \textit{Swift} observations concurrent with \emph{Fermi}-LAT detections. In all cases for the classical novae, the first X-ray detection only occurs after the $\gamma$-ray emission falls below the sensitivity limit of \textit{Fermi}-LAT (Figure~\ref{fig:gamma schwarz}).

\par The simplest explanation for the \emph{Swift}-XRT non-detections during $\gamma$-rays is that the shocks are deeply embedded within the nova ejecta due to their high density (among the highest for astrophysical events: $\sim10^{10}$\,cm$^{-3}$; see figure 1 in \citealt{Metzger_etal_2016}). Such high densities imply substantial absorbing columns ahead of the shocks, which can absorb photons with energies $\lesssim$10\,keV. 
The other explanation for the X-ray non-detections is that the thermal energy of the shocked material is sapped by cold regions around the shocks before it can be radiated, implying a suppression of the shock's temperature, i.e., the shocks do not reach X-ray energies \citep{Steinberg_Metzger_2018}. This would lead to a suppression of the X-ray emission that can be detected by \textit{Swift}. 

\par
\begin{figure}
    \centering
    \includegraphics[width=1.08\columnwidth]{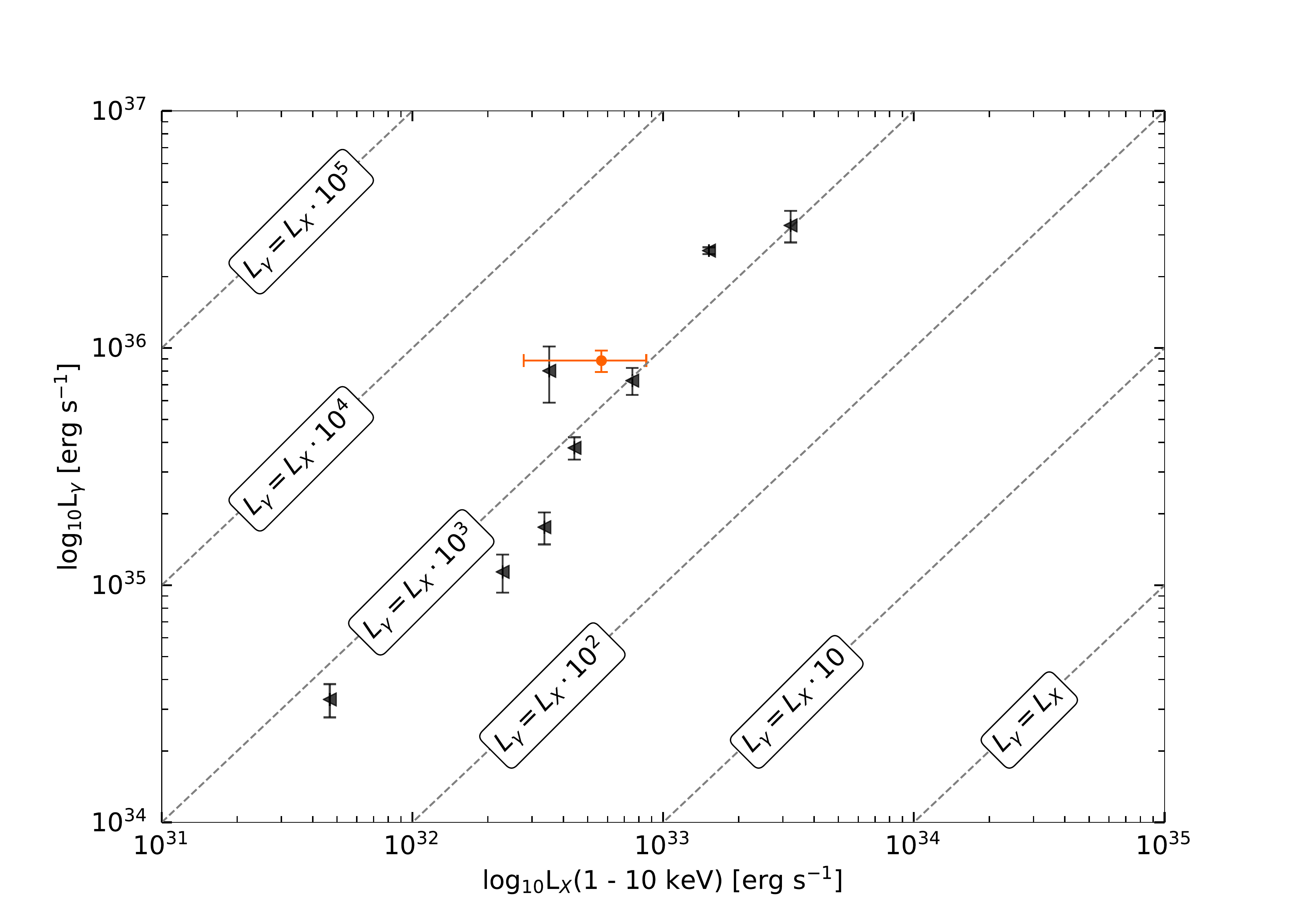}
    \caption{Comparison of hard X-ray and $\gamma$-ray luminosities, for novae with concurrent X-ray and $\gamma$-ray data. X-ray luminosities are in the 1--10 keV band and corrected for Galactic absorption (as given in Table \ref{table:characteristics}) and represent the faintest (and most constraining) X-ray points during the $\gamma$-ray detection period.  Novae represented as black triangles denote that they are non-detections in the hard X-ray band, and 3$\sigma$ upper limits are plotted. V407 Cyg is represented as an orange circle, as the only nova that had a $\geq$3$\sigma$ \textit{Swift} detection during the $\gamma$-ray detection period. $\gamma$-ray luminosities are calculated in the 100 MeV--300 GeV band using parameters listed in Table \ref{table:gamma}. Dashed lines guide the eye for estimating $L_{\gamma}/L_X$.}
    \label{fig:gamma_xray_flux}
\end{figure}

 To constrain the conditions in nova shocks, we compare the \emph{Swift}-XRT upper limits on the 1--10 keV X-ray luminosity with concurrent GeV $\gamma$-ray luminosities from \emph{Fermi}-LAT (Figure \ref{fig:gamma_xray_flux}). We convert the time-averaged \emph{Fermi}-LAT $\gamma$-ray count rates listed in Table \ref{table:gamma} to $\gamma$-ray fluxes assuming a single power-law spectrum and  photon indices also listed in Table \ref{table:gamma}. The fluxes are then converted to luminosites over the energy range 100 MeV--300 GeV, assuming the distances in Table \ref{table:characteristics}. The resulting $\gamma$-ray luminosities span a few $\times 10^{34}$ to a few $\times 10^{36}$ erg s$^{-1}$. The X-ray luminosities are as estimated for Figure \ref{fig:hard_luminosity} (\S \ref{subsec:hard_lum}).
V392 Per, V549 Vel, V407 Lup, and V959 Mon are not plotted in Figure \ref{fig:gamma_xray_flux}, as there were no X-ray observations during their $\gamma$-ray emitting periods. The $\gamma$-ray luminosities are factors at least $10^{2}$ to $10^{4}$ times more luminous than the X-ray upper limits,  with most of the novae clustered around $L_{\gamma}/L_X \approx 10^{3}$. The ratios show a remarkable correlation, but no strong conclusions should be made as the plot is comparing a single data point in the X-rays per each nova to an averaged $\gamma$-ray luminosity over the detection period. The \emph{Swift} upper limit depth is heavily dependent on exposure time and the background count rate. 

\par Motivated by \textit{Swift} non-detections,  
researchers have begun searching for even harder X-rays during the $\gamma$-ray bright phase using the \textit{NuSTAR} satellite \citep{2013ApJ...770..103H}, with instruments on board sensitive to photons with energies up to 79 keV. While softer X-rays are absorbed, harder X-rays $>$10 keV are expected to escape the dense ejecta, even in the early days of the eruption, due to the decreasing bound-free cross-section at high photon energies \citep{2015MNRAS.450.2739M}.
Harder X-rays have now been detected with \emph{NuSTAR} from three classical novae concurrently with $\gamma$-rays, namely V5855 Sgr, V906 Car, and YZ Ret (YZ Ret is not included in our sample as it erupted in 2020; \citealt{Nelson_etal19, 2020MNRAS.497.2569S, Sokolovsky+20_ret20}).
Spectral analysis of these \emph{NuSTAR} data show low-luminosity hard X-ray emission ($\sim 10^{33}-10^{34}$ erg~s$^{-1}$) originating from hot plasma ($kT \approx 5-10$ keV) and absorbed by large column densities (N(H) $\approx 10^{23}-10^{24}$ cm$^{-2}$).

\par Even with these \emph{NuSTAR} detections corrected for internal absorption, the $L_{\gamma}/L_X$ ratio is still $\gtrsim 10-10^{2}$. The high $L_{\gamma}/L_X$ observed with both \textit{Swift} and \textit{NuSTAR} is surprising because only a fraction of the shock's power should be going into producing $\gamma$-rays given the predicted efficiency for particle acceleration \citep{2015MNRAS.450.2739M}. Meanwhile, the high post-shock densities imply that the shocks should be radiative, and so the majority of the shock luminosity should be promptly transferred to radiative luminosity, which is naively expected to emerge in the X-ray band \citep{2015MNRAS.450.2739M, Li_etal_2017, 2020NatAs...4..776A}. 
\citet{Nelson_etal19} and \citet{2020MNRAS.497.2569S} propose several scenarios that could yield a much higher $\gamma$-ray luminosity compared to X-rays, including separate shocks producing the X-rays and $\gamma$-rays, suppression of the X-rays by corrugated shock fronts \citep{Steinberg_Metzger_2018}, remarkably efficient particle acceleration, or that modeling the shocks as radiative is an improper assumption.

\par Of the 12 classical novae investigated here, 10 eventually show 1--10 keV hard X-ray emission detectable by \emph{Swift}-XRT.
This late emergence of the hard X-ray emission can be partially explained by a drop in the density of the ejecta as they expand---leading to a decrease in the absorbing column ahead of the shocks. But the faint \emph{NuSTAR} detections imply that it is not only large absorbing columns that are leading to \emph{Swift} non-detections; the X-ray luminosity is also intrinsically low.

\subsection{The exception: novae with giant companions are detected in X-rays concurrent with $\gamma$-rays}\label{sec:v407}


\par Although evolved giant companions are relatively rare in nova-hosting binaries, the first-ever nova detected by \emph{Fermi}-LAT, V407~Cyg, was accompanied by a Mira giant donor \citep{2010Sci...329..817A}. Previous to the nova eruption in 2010, V407~Cyg was well known as a D-type symbiotic star \citep[e.g.,][]{Munari_etal90, Kolotilov_etal98, Kolotilov_etal03}. The giant donor's wind was dense, with a mass-loss rate of $\sim 10^{-6}$ M$_{\odot}$ yr$^{-1}$, resulting in a rich circumbinary medium \citep{Chomiuk_etal_2012}.

\par During its 2010 nova eruption (discovered on 2010 March 10), V407~Cyg displayed faint but detectable X-rays in the first \emph{Swift}-XRT observations of the nova eruption (four days after nova discovery; Figure~\ref{fig:V407_Cyg}, \citealt{Shore_etal11, Nelson_etal_2012}). Over the next $\sim$20 days following the nova discovery, the X-ray flux rapidly brightened by a factor of $\sim$10. During this same time period, V407~Cyg was detected as a GeV $\gamma$-ray source by \emph{Fermi}-LAT  \citep{2010Sci...329..817A}. V407~Cyg is the only nova in our sample with concurrent \emph{Swift}-XRT and \emph{Fermi}-LAT detections.

\par Both the X-rays and the $\gamma$-rays in V407~Cyg are attributed to the interaction of the nova ejecta with the circumbinary medium \citep{Orlando_Drake12, Martin_Dubus13}. The X-ray flux rises in the first three weeks because the absorbing column might have dropped, while the X-ray emission measure grows. The absorbing column, even at early times, is never much higher than $N(H) \approx 10^{23}$ cm$^{-2}$ \citep{Nelson_etal_2012}. This can be contrasted with the absorbing columns of $\gtrsim {\rm few} \times 10^{23}$ cm$^{-2}$ for the internal shocks observed in classical novae with dwarf companions (e.g., \citealt{Nelson_etal19, 2020MNRAS.497.2569S}). Therefore, V407~Cyg hints that X-rays can be detected concurrently with $\gamma$-rays if the nova drives {\it external} shocks (i.e., interaction with pre-existing circumbinary material), as opposed to more deeply-absorbed shocks internal to the nova ejecta.

\par This hypothesis is supported by two additional novae with giant companions that were marginally detected by \emph{Fermi}-LAT between 2010 and 2018: V745~Sco and V1535~Sco \citep{Franckowiak_etal_2018}. Hints of $\gamma$-ray emission from V745~Sco were obtained at 2--3$\sigma$ significance in the first two days of its 2014 nova eruption \citep{2014ATel.5879....1C}. Bright hard X-ray emission was also observed during this time, with $N(H) = {\rm few} \times 10^{22}$ cm$^{-2}$ (\citealt{Delgado_Hernanz19}; again, substantially lower than the absorbing columns observed for shocks in classical novae). Similarly, V1535~Sco was marginally detected in $\gamma$-rays during the first seven days of its 2015 eruption \citep{Franckowiak_etal_2018}, and hard X-rays were concurrently detected by \emph{Swift}-XRT (on day 4; \citealt{Linford_etal17}). Although these $\gamma$-ray detections are marginal, they support a scenario where nova shocks with external circumbinary material (as occur in binaries with giant companions) are characterized by lower density, less embedded environments, in comparison with shocks that occur internal to nova ejecta in binaries with dwarf companions.

It is worth noting that the high $L_{\gamma}/L_X$ observed in V407~Cyg (see Figure~\ref{fig:gamma_xray_flux}) could not be explained by high absorption or X-ray suppression in this case, given the less embedded environments of the shocks. However, a detailed analysis of the shocks in novae with evolved secondaries is outside the scope of this paper and will be the topic of future projects.

\subsection{Why are some $\gamma$-ray detected novae never detected in X-rays?} \label{no_detection_novae}

\par Out of the 13 novae in our sample, only two were never detected as X-ray sources with \textit{Swift}, namely V1324~Sco (Figure~\ref{fig:V1324_Sco}; \citealt{Finzell_etal_2018}) and V5856~Sgr\footnote{After further analysis of the WT data of V5856~Sgr, there is a possible X-ray detection on day 149, but the online generator did not find any detection. This is mainly affected by the estimate of background contribution for faint objects observed in WT mode.} (Figure~\ref{fig:V5856_Sgr}; \citealt{Li_etal_2017}). V5856~Sgr had only two \textit{Swift} observations (15 and 149 days after discovery) which makes it difficult to draw conclusions about this nova as its X-ray emission could have been missed (as we might have missed the X-ray emission from e.g., V357~Mus if observations of it had been similarly sparse). However, V1324~Sco was followed with \textit{Swift} between days 30 and $\sim$ 500 after eruption and was still never detected. 

\par There are a few reasons that might explain why V1324~Sco was not detected: lack of correlation between $\gamma$-ray luminosity and X-ray luminosity, distance, and/or absorption. While V1324 Sco was not detected in X-rays, it is among the brightest novae detected in $\gamma$-rays. If X-ray luminosity does not scale with $\gamma$-ray luminosity, this could explain the difference between the two.



\par Distance, however, appears to be an important factor to the detection of X-rays. V1324 Sco is the farthest nova of our sample ($\gtrsim$6.5 kpc), and Figure \ref{fig:hard_luminosity} shows that this translates to less sensitive upper limits on the hard X-ray luminosity.
We compared the flux of each nova's first 1--10 keV detection to what it would be at V1324 Sco's distance (also correcting for the additional interstellar absorption). 
This analysis revealed that five novae would have been non-detections at the distance and $N(H)$ of V1324~Sco: V906 Car, V357 Mus, V5856 Sgr, V5855 Sgr, and V5668 Sgr. We therefore conclude that distance is probably the reason why V1324 Sco was not detected by \emph{Swift}-XRT.



\begin{figure*}[t]
    \centering
    \includegraphics[width=\textwidth]{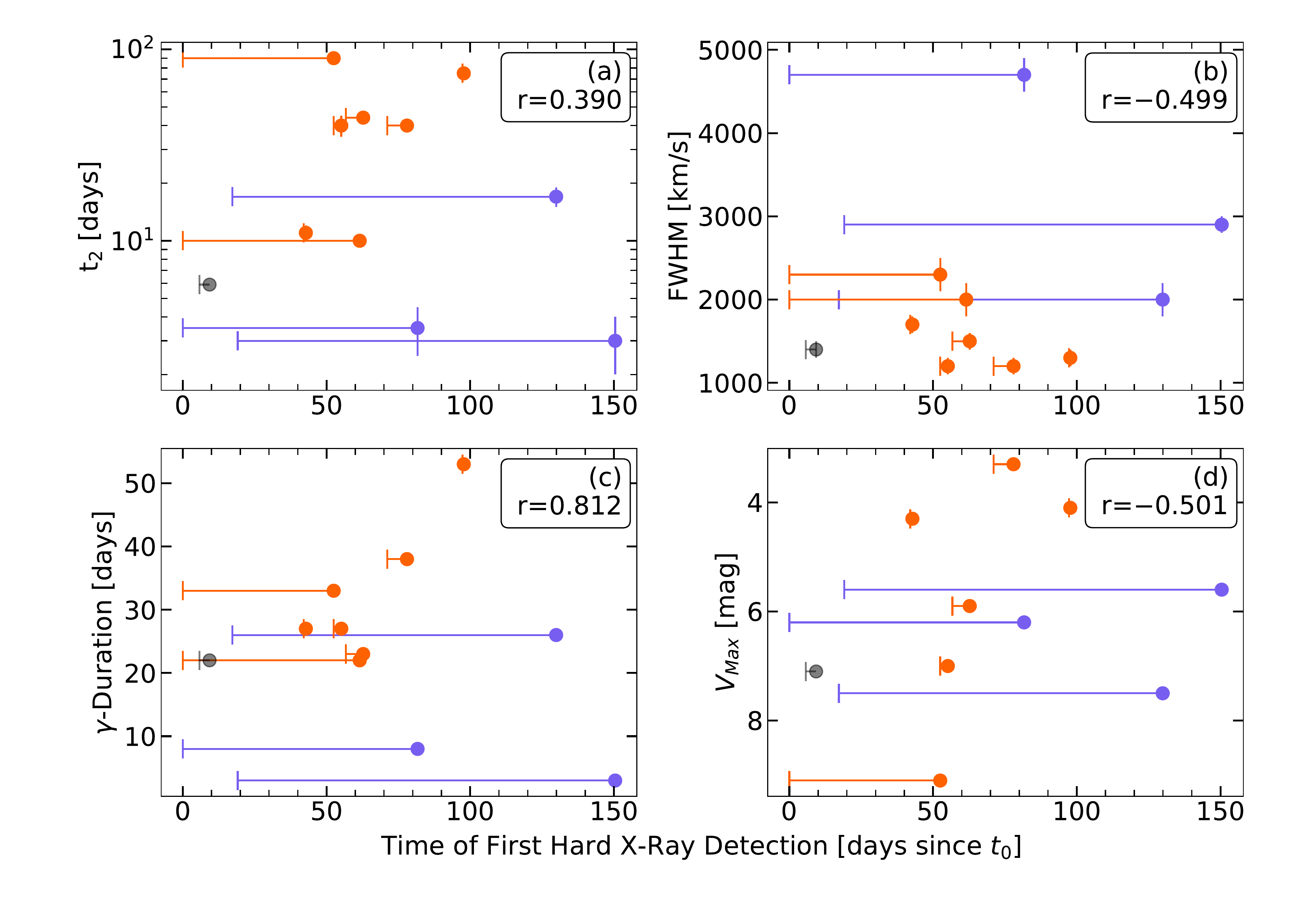}
    \caption{The date of the first hard X-ray detection since $t_0$ for each nova plotted against $t_2$ (panel a), FWHM (measured from the Balmer lines a few days after optical maximum; panel b), duration of detectable $\gamma$-rays (panel c), and apparent magnitude at optical peak ($V_{\mathrm{max}}$; panel d) for the novae in our sample for which a measurement is available. The error in the first hard X-ray detection represents a lower limit on this quantity, extending to the date of the last non-detection; if the arrows extend to $t=t_0$, the first \textit{Swift}-XRT observation was a 3$\sigma$ detection. V407 Cyg is represented in dark gray in each panel as it has a giant companion (\S \ref{sec:v407}).  The $r$ value denoted in the upper right corner of each panel is the Pearson correlation coefficient derived by using the grey and orange circles in each panel and excluding any weight for the error bars. V392 Per, V407 Lup, and V5855~Sgr are plotted in purple as the actual hard X-ray start for these novae is uncertain (see \S\ref{subsec:trends} for more details), therefore we exclude them from the correlation fitting. V1324~Sco and V5856~Sgr were never detected in X-rays by \textit{Swift} (see \S\ref{no_detection_novae}) so they do not appear in the plots; V959 Mon does not appear in panel d as a maximum V-band magnitude could not be determined.}
    \label{fig:trends}
\end{figure*}

\subsection{What determines when the hard X-rays appear?}
\label{subsec:trends}

\par Part of the intention of this project was to study a sample of $\gamma$-ray detected novae in order to analyze possible trends in the data. 
In Figure~\ref{fig:trends}, we plot the date of the first hard X-ray detection against other nova properties described in \S \ref{sec:gamma} and \S\ref{sec:optical}: $t_2$, FWHM of Balmer emission lines (after optical peak), apparent magnitude at optical peak ($V_{\mathrm{max}}$), and the duration of the $\gamma$-ray detection to check for any correlations between these parameters. The Pearson correlation coefficient is shown in the top right corner of each panel. 

\par Since the timing of \emph{Swift} observations are different for each nova, it is challenging to draw conclusions about correlations between these parameters. For novae with extremely bright supersoft emission, it is possible that the harder shock component is contaminated by the supersoft component (see \S\ref{subsec:hard_lum} for more discussion). The cadence of novae V407~Lup, V392~Per, and V5855~Sgr was interrupted by solar conjunction and observation schedules. In addition, V407~Lup and V5855~Sgr were first detected during a bright supersoft phase which caused large uncertainties on the first hard X-ray start date for these novae as plotted in Figure~\ref{fig:trends}. Because of these complications, we exclude these novae from the fitting done to derive the correlation coefficients.

\par Based on panel (a) in Figure~\ref{fig:trends}, visual inspection indicates earlier hard X-ray emission for faster novae (characterized by smaller $t_2$)---particularly for novae with extensive \emph{Swift} follow-up (with short error bars in Figure~\ref{fig:trends}). The Pearson coefficient factor of $r = 0.39$ derived for novae with higher-quality data also implies that there may be a weak correlation. Interestingly, we find a weak anti-correlation between the time of first hard X-ray detection and FWHM ($r= -0.50$  shown in panel (b).
A nova characterized by a faster optical light curve (short $t_2$) should typically have higher ejecta expansion velocities (large FWHM; \citealt{Shafter_etal_2011}). In this case, the ejecta are expected to expand, drop in density, and become optically thin to the X-ray emitting shocked regions more rapidly than slower novae. There are hints that we may be observing these trends in Figure \ref{fig:trends}, but a larger sample of novae will need to be observed in the future in order to confirm these hints.

\par In panel (c), a Pearson correlation coefficient of 0.81 implies a likely correlation between the duration of the $\gamma$-ray detection and the first hard X-ray detection. To first order, this is expected given that none of the novae in our sample recorded \emph{Swift} X-ray detections concurrent with the $\gamma$-ray emission and were only detected after this period ended. But this correlation may hold important clues as to the drivers of shocks in novae, as the only other quantity that has been observed to potentially correlate with $\gamma$-ray duration is $\gamma$-ray fluence \citep{2016ApJ...826..142C, Franckowiak_etal_2018}.

Panel (d) shows some indication of a correlation between the peak brightness of the nova and the time of first hard X-ray detection (note that an anti-correlation here is a correlation with brightness, due to the ``flipped" magnitude scale). However, again this correlation is weak and requires a larger sample or higher cadence data to test. 

\par In summary, although there are intriguing hints at correlations, it is challenging to draw  conclusions from the current sample---the number of novae detected in $\gamma$-rays with dedicated multi-wavelength follow up is still small. Additional  novae with high-cadence \textit{Swift}-XRT and optical follow-up added to the current sample will allow us to draw better conclusions in the future.

\section{Summary and conclusions}
\label{sec:conc}

\par We have investigated the hard (1--10 keV) X-ray emission of 13 $\gamma$-ray emitting novae using \textit{Swift}-XRT. Novae have long been observed to emit X-rays from hot ($kT \approx$ 1--10 keV)  optically-thin plasma, presumably from shocked gas \citep{1994MNRAS.271..155O, 2008ApJ...677.1248M}. The  \textit{Swift}-XRT light curves show evidence of hard X-ray emission from shocks in at least 7 out of the 13 novae studied, typically peaking several months after the start of eruption with luminosities $\sim 10^{33}-10^{34}$ erg~s$^{-1}$.

However, of the 9 novae with \textit{Swift}-XRT observations during the $\gamma$-ray detection phase (typically a few weeks around optical maximum), eight yielded X-ray non-detections during these early times.
The only nova showing X-ray emission concurrently with a \textit{Fermi} $\gamma$-ray detection is V407~Cyg, which has a giant secondary. We suggest that the non-detection of early X-ray emission from the other eight novae (all with dwarf companions) is due to a combination of large column densities ahead of the shocks absorbing the X-rays, and X-ray suppression by corrugated shock fronts \citep[e.g.,][]{2015MNRAS.450.2739M,Steinberg_Metzger_2018}. The early X-ray detection of V407~Cyg (and possibly other novae with evolved companions) confirms that the shocks in symbiotic systems are external (between the nova ejecta and circumbinary material), rather than internal to the nova ejecta as claimed for novae with dwarf companions.


\par As more $\gamma$-ray emitting novae are discovered and followed up at other wavelengths, we will be able to better constrain the physical parameters of the shocks and further investigate the conditions of their surrounding media.

\section*{Acknowledgments}

We are grateful to Tommy Nelson and Brian Metzger for conversations that inspired this work.
ACG, EA, LC, KVS, and JS are grateful for the support of NASA \emph{Fermi} grant 80NSSC18K1746, \emph{NuSTAR} grant 80NSSC19K0522, NSF award AST-1751874, and a Cottrell Scholarship of the Research Corporation. KLP acknowledges funding from the UK Space Agency.
KLL is supported by the Ministry of Science and Technology of the Republic of China (Taiwan) through grants 108-2112-M-007-025-MY3 and 109-2636-M-006-017, and he is a Yushan (Young) Scholar of the Ministry of Education of the Republic of China (Taiwan).

This work made use of data supplied by the UK Swift Science Data Centre at the University of Leicester. We acknowledge with thanks the variable star observations from the AAVSO International Database contributed by observers worldwide and used in this research. We also acknowledge with thanks the Astronomical Ring for Access to Spectroscopy ARAS observers for their optical spectroscopic observations

\bibliography{nova_biblio}







\clearpage

\appendix

\renewcommand\thetable{\thesection.\arabic{table}}    
\renewcommand\thefigure{\thesection.\arabic{figure}}   
\setcounter{figure}{0}
\setcounter{table}{0}


\section{\textit{Swift} X-ray Light Curves}
\label{appB}
In this Appendix, we present the XRT X-ray (0.3--10\,keV) light curves for all the novae in our sample. We also plot the soft (0.3--1.0\,keV) and hard (1.0--10\,keV) light curves.

\begin{figure*}[h!]
    \centering
    \includegraphics[width=0.8\textwidth]{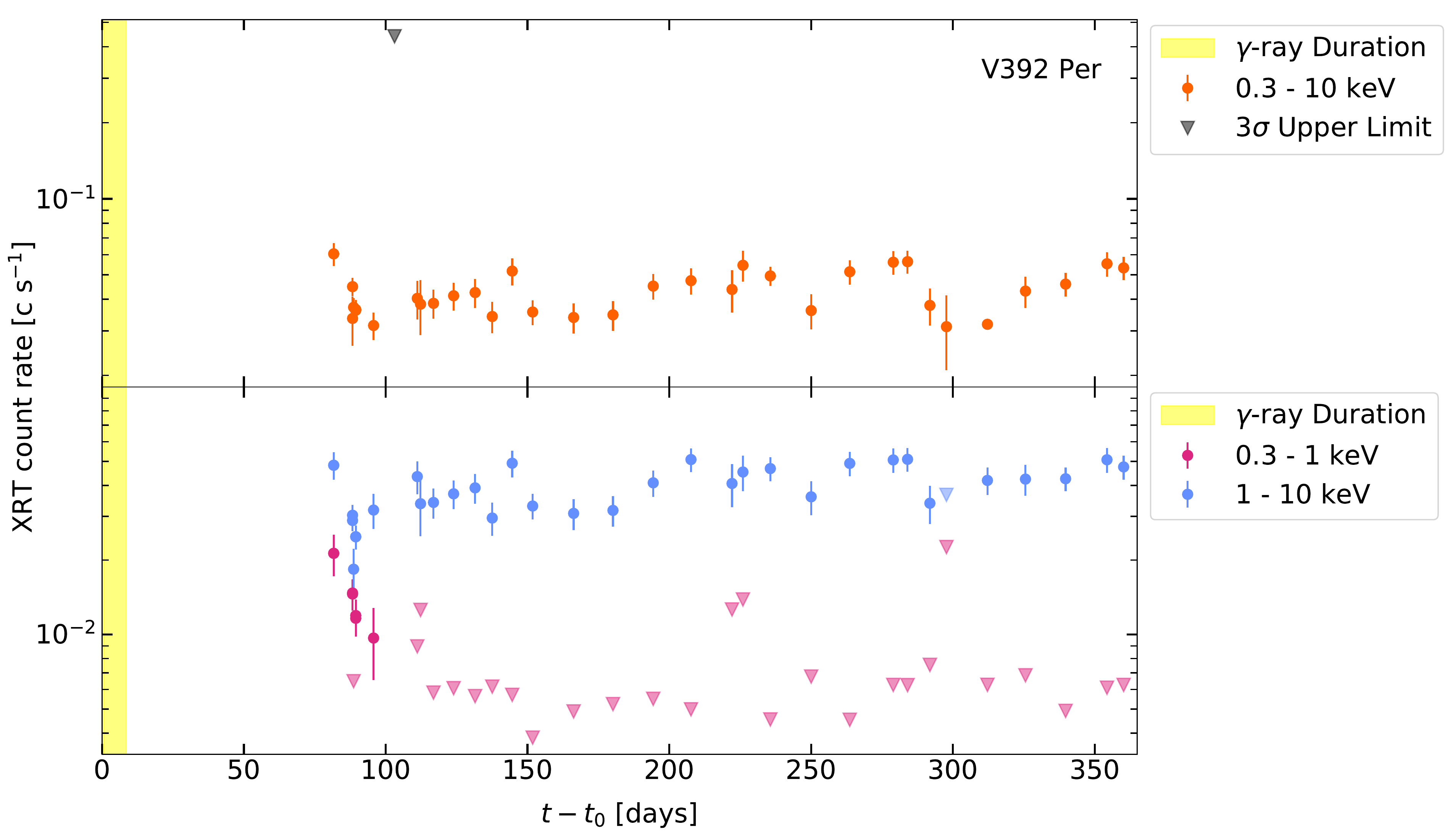}
    \caption{\textit{Swift}-XRT X-ray light curves of V392~Per, plotted as a function of days since discovery ($t-t_0$). The upper panel plots the count rate over the full 0.3\,--\,10\,keV energy range, with orange points representing detections and dark grey triangles representing 3$\sigma$ upper limits. The lower panel splits the counts into hard (1.0\,--\,10\,keV) and soft (0.3\,--\,1.0\,keV) X-ray bands. The soft band is plotted in magenta, and the hard band is plotted in blue; in both cases, circles represent detections and triangles represent 3$\sigma$ upper limits. The time range wherein \emph{Fermi}-LAT detected $\gamma$-rays with $>$3$\sigma$ significances is marked as a yellow bar. Any purple points come from the hard and soft points overlapping each other. Similarly, darker colored full-band upper limits come from overlapping points.}
    \label{fig:V392_Per}
\end{figure*}

\begin{figure*}
    \centering
    \includegraphics[width=0.8\textwidth]{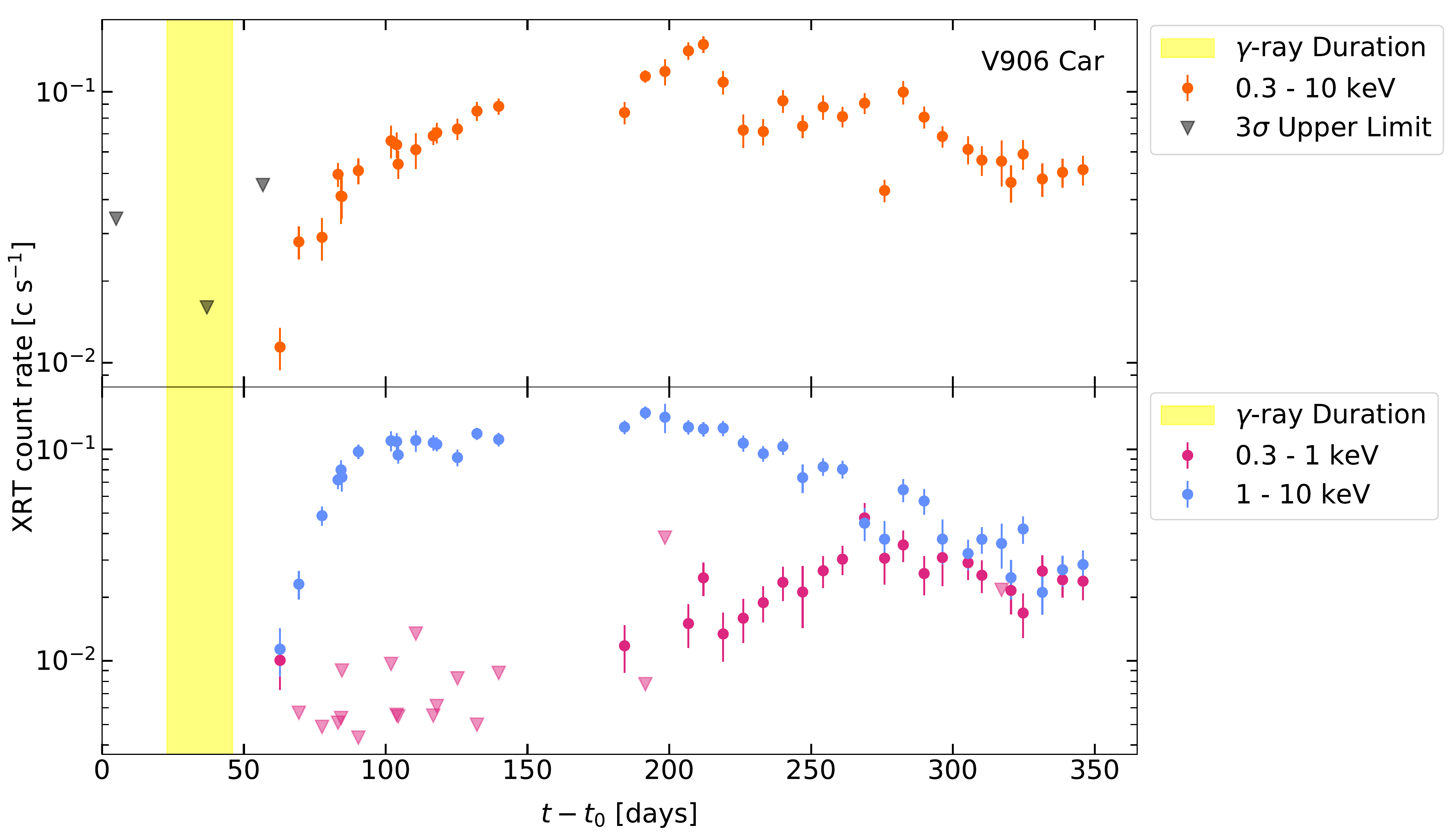}
    \caption{\textit{Swift}-XRT X-ray light curves of V906~Car (ASASSN-18fv). See Figure \ref{fig:V392_Per} for more details.}
    \label{fig:V906_Car}
\end{figure*}

\begin{figure*}[!t]
    \centering
    \includegraphics[width=0.8\textwidth]{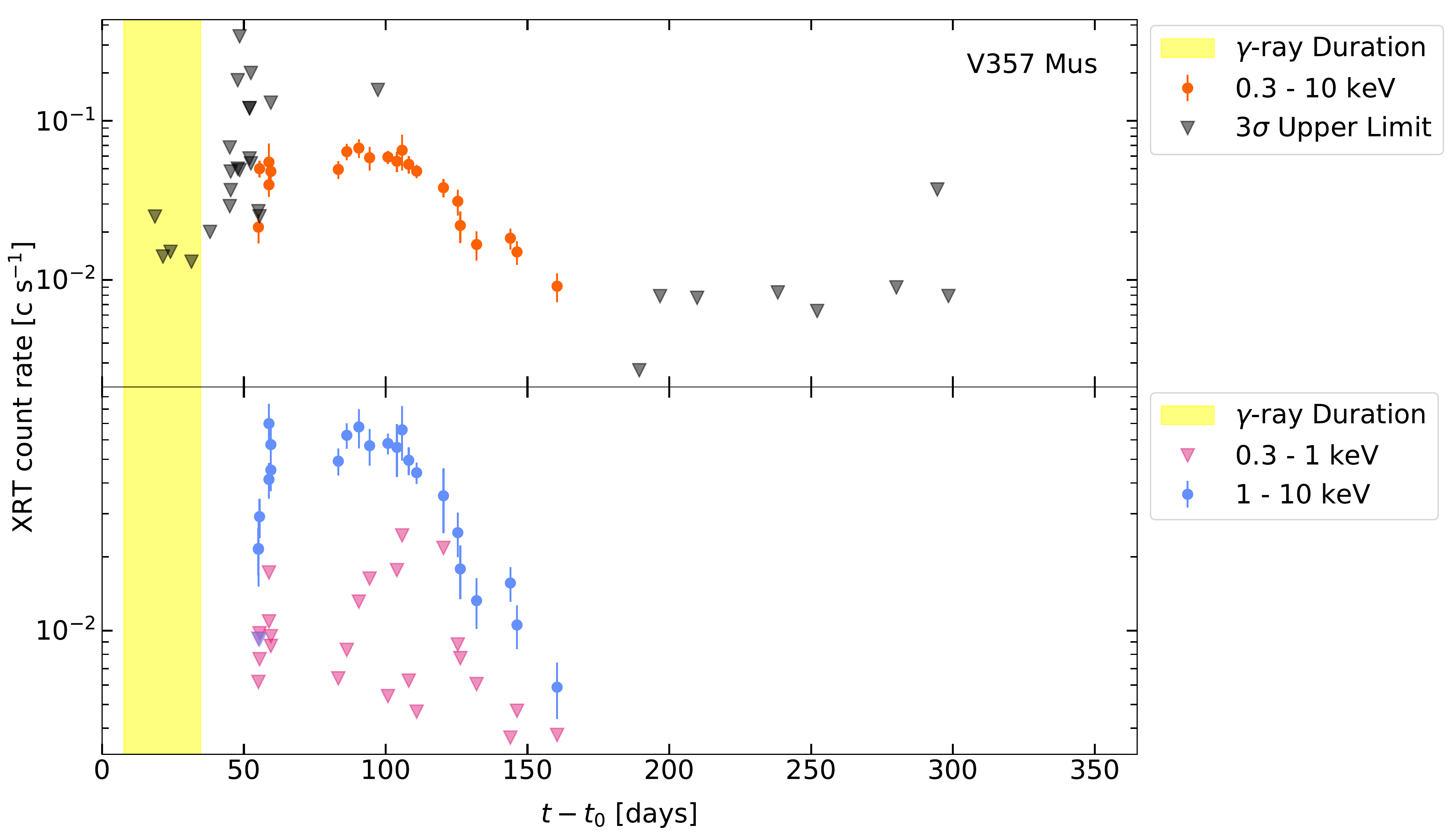}
    \caption{\textit{Swift}-XRT X-ray light curves of V357~Mus. See Figure \ref{fig:V392_Per} for more details.}
    \label{fig:V357_Mus}
\end{figure*}

\begin{figure*}[!t]
    \centering
    \includegraphics[width=0.8\textwidth]{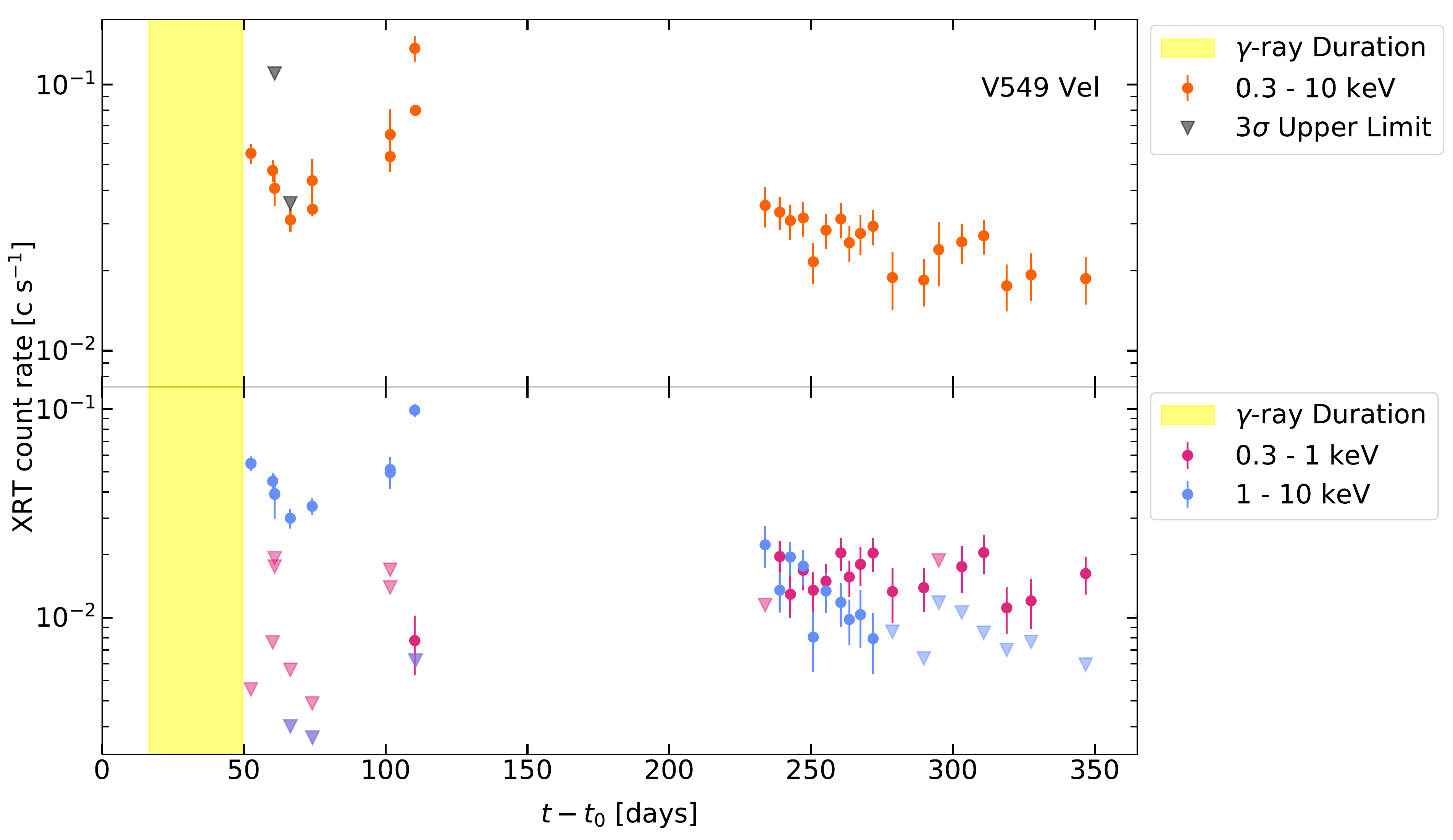}
    \caption{\textit{Swift} XRT X-ray light curves of V549~Vel (ASASSN-17mt). See Figure \ref{fig:V392_Per} for more details. }
    \label{fig:V549 Vel}
\end{figure*}

\begin{figure*}[!t]
    \centering
    \includegraphics[width=0.8\textwidth]{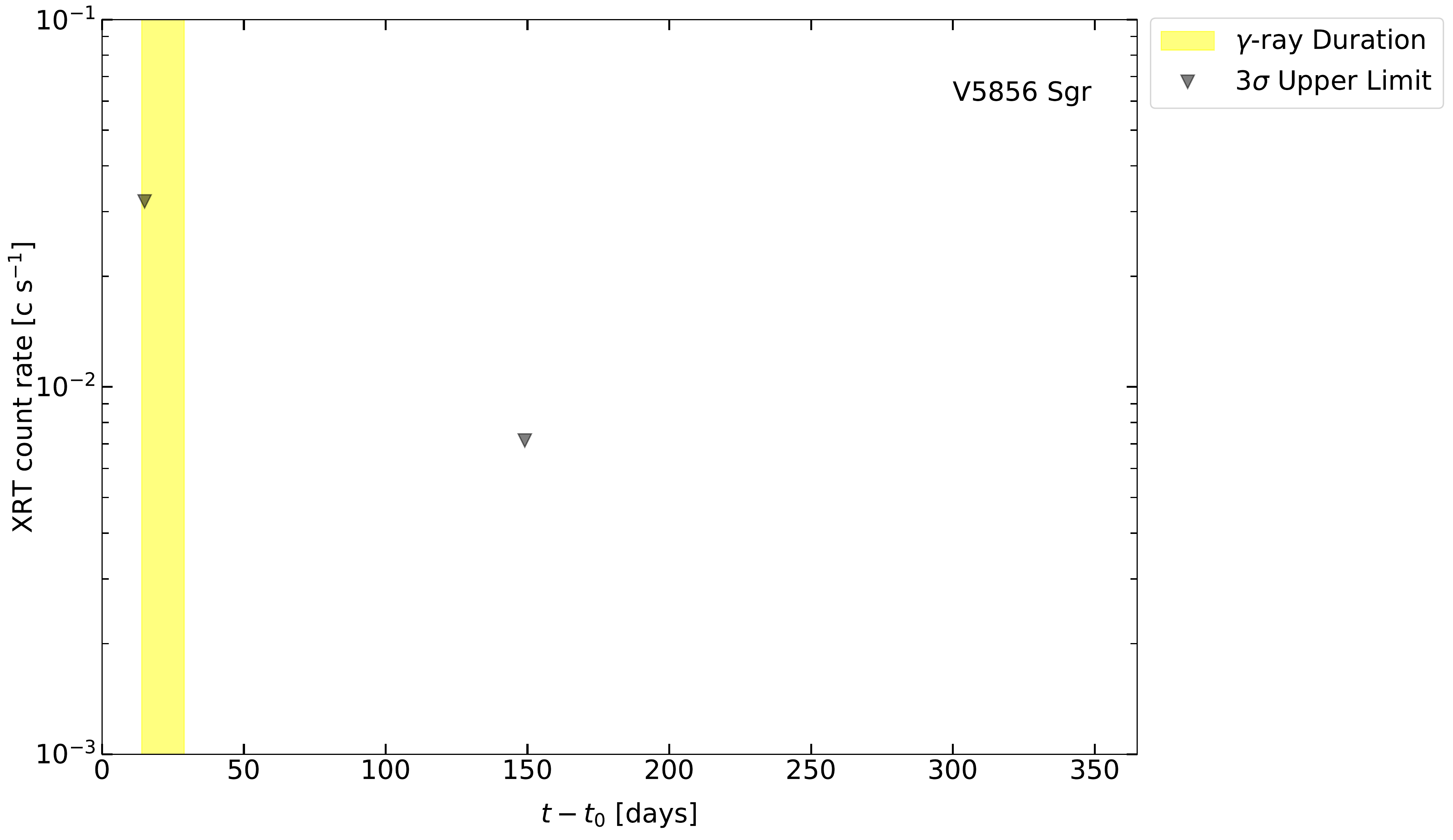}
    \caption{\textit{Swift}-XRT X-ray light curve of V5856~Sgr (ASASSN-16ma). See Figure \ref{fig:V392_Per} for more details.}
    \label{fig:V5856_Sgr}
\end{figure*}

\begin{figure*}[!t]
    \centering
    \includegraphics[width=0.8\textwidth]{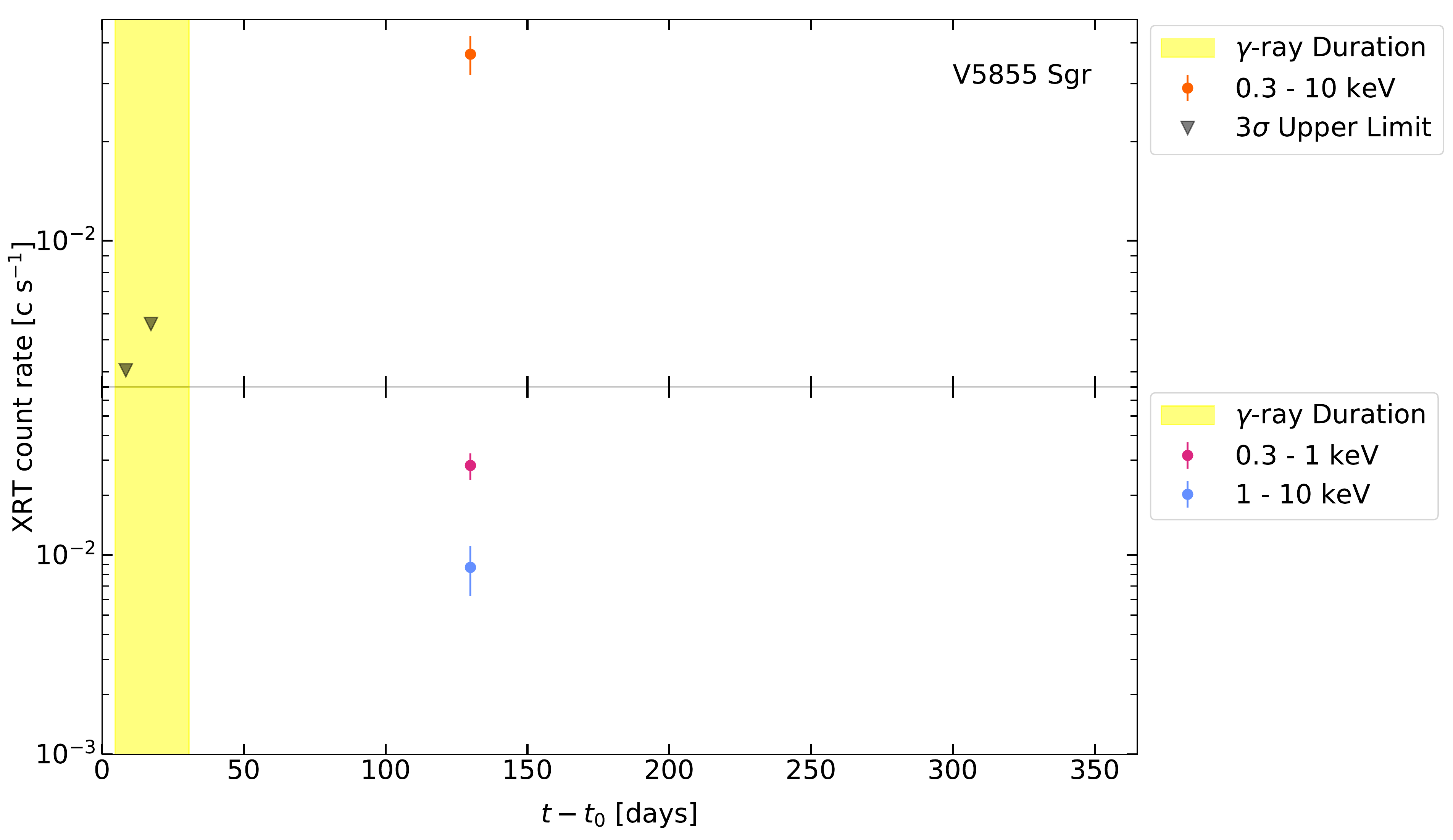}
    \caption{\textit{Swift}-XRT X-ray light curves of V5855~Sgr. See Figure \ref{fig:V392_Per} for more details.}
    \label{fig:V5855_Sgr}
\end{figure*}

\begin{figure*}[!t]
    \centering
    \includegraphics[width=0.8\textwidth]{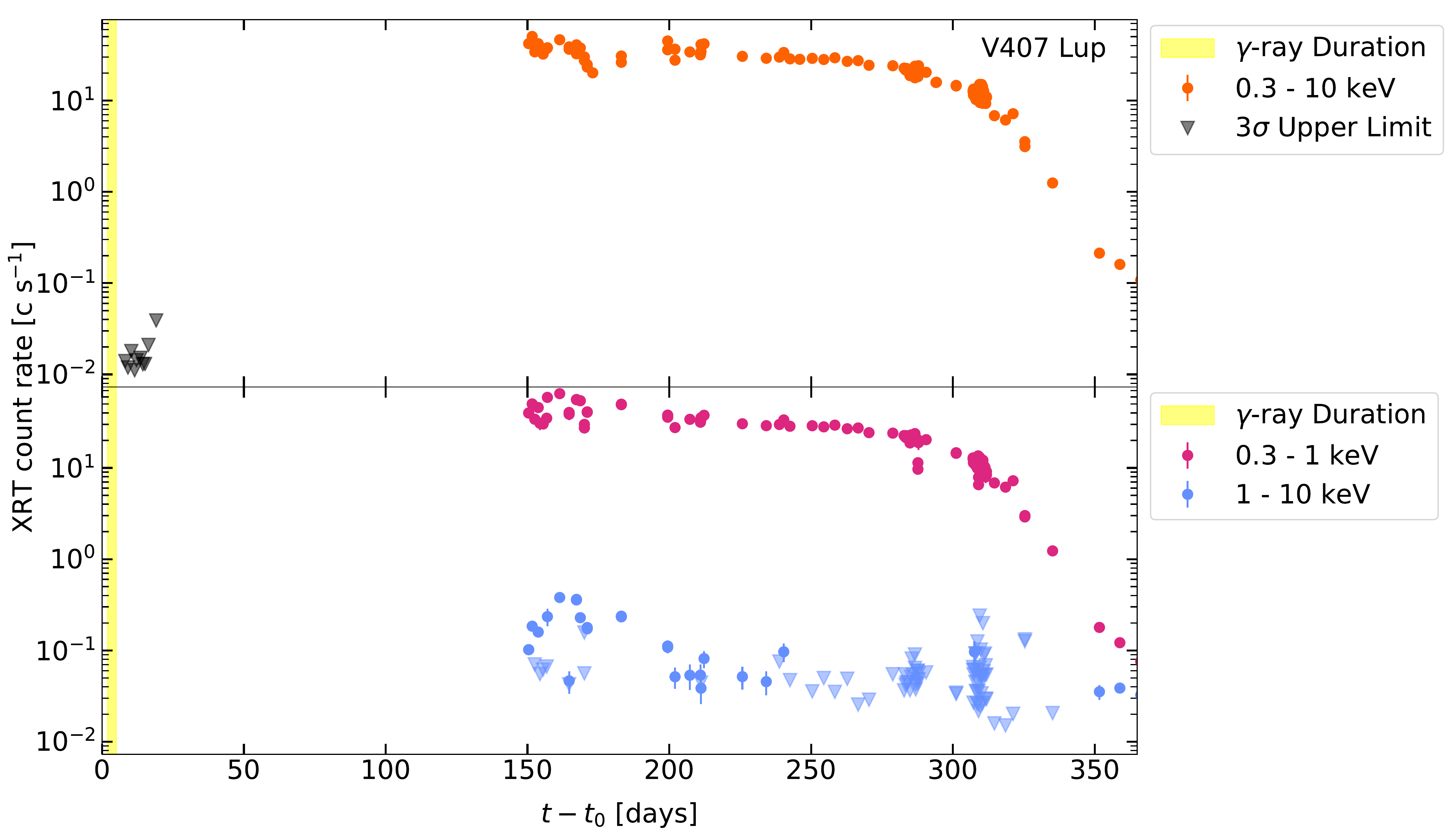}
    \caption{\textit{Swift}-XRT X-ray light curves of V407~Lup (ASASSN-16kt). See Figure \ref{fig:V392_Per} for more details.}
    \label{fig:V407_Lup}
\end{figure*}

\begin{figure*}[!b]
    \centering
    \includegraphics[width=0.8\textwidth]{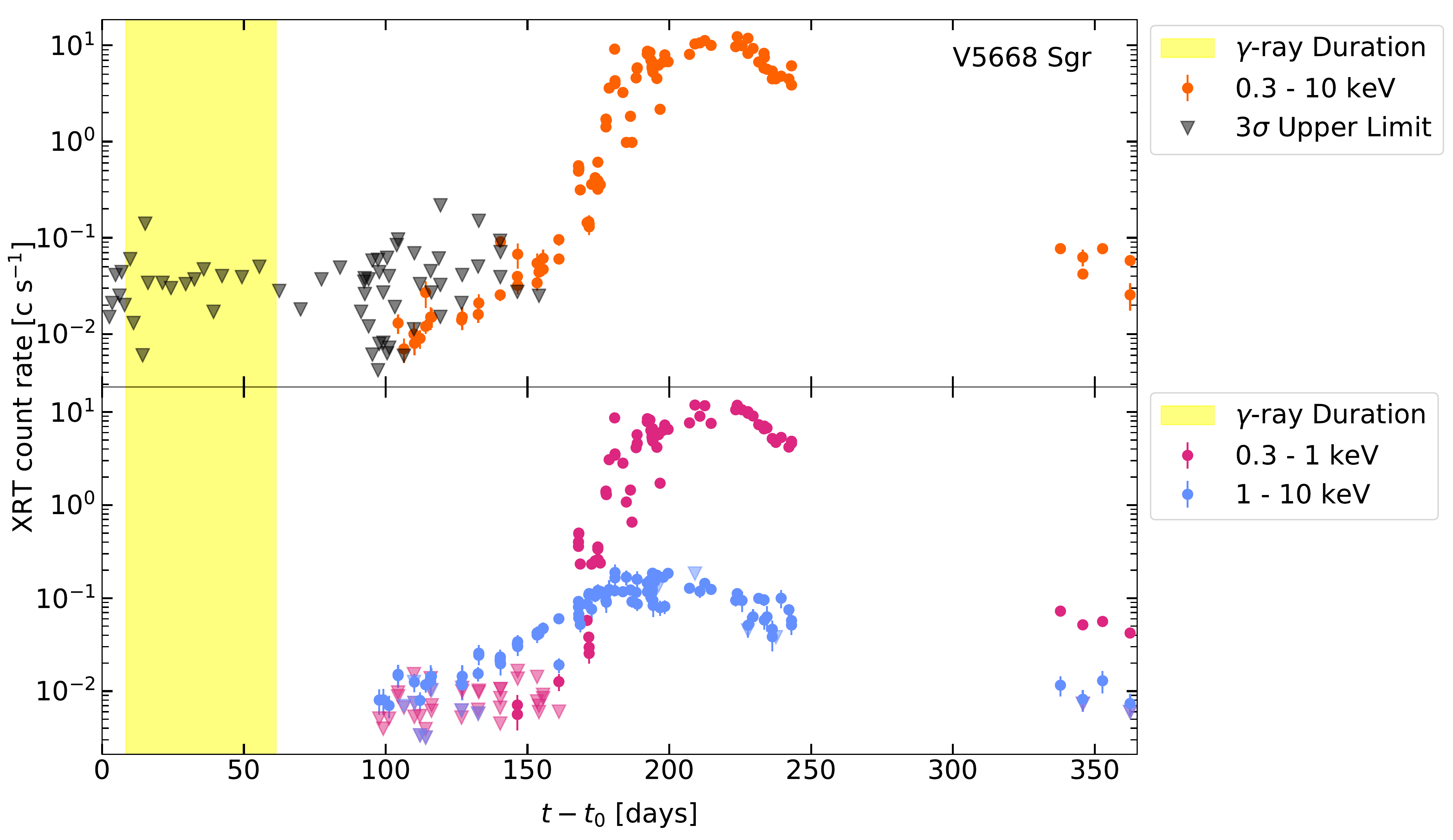}
    \caption{\textit{Swift}-XRT X-ray light curves of V5668~Sgr. See Figure \ref{fig:V392_Per} for more details.}
    \label{fig:V5568_Sgr}
\end{figure*}

\begin{figure*}[!b]
    \centering
    \includegraphics[width=0.8\textwidth]{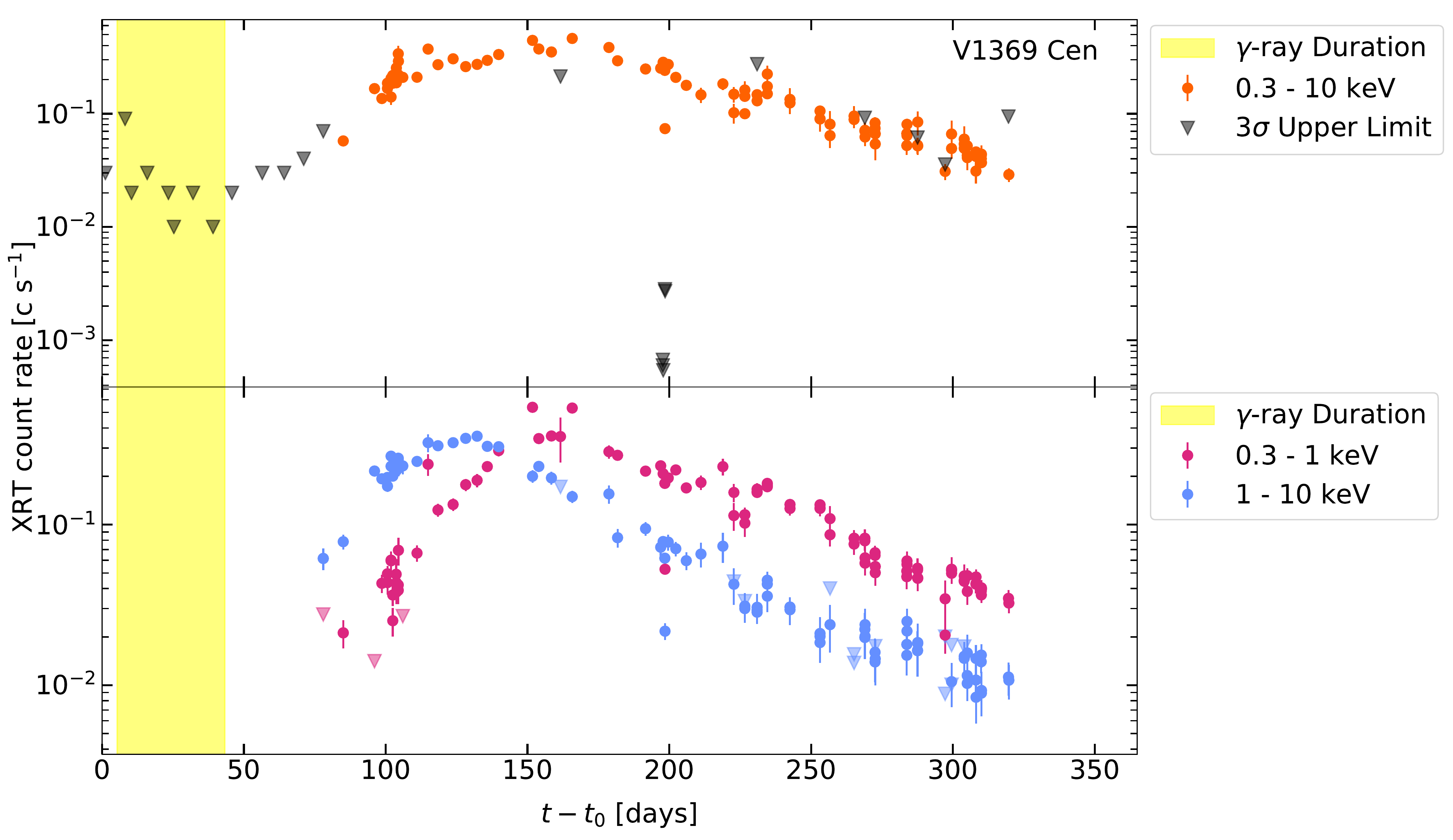}
    \caption{\textit{Swift}-XRT X-ray light curves of V1369~Cen. See Figure \ref{fig:V392_Per} for more details.}
    \label{fig:V1369_Cen}
\end{figure*}

\begin{figure*}[!b]
    \centering
    \includegraphics[width=0.8\textwidth]{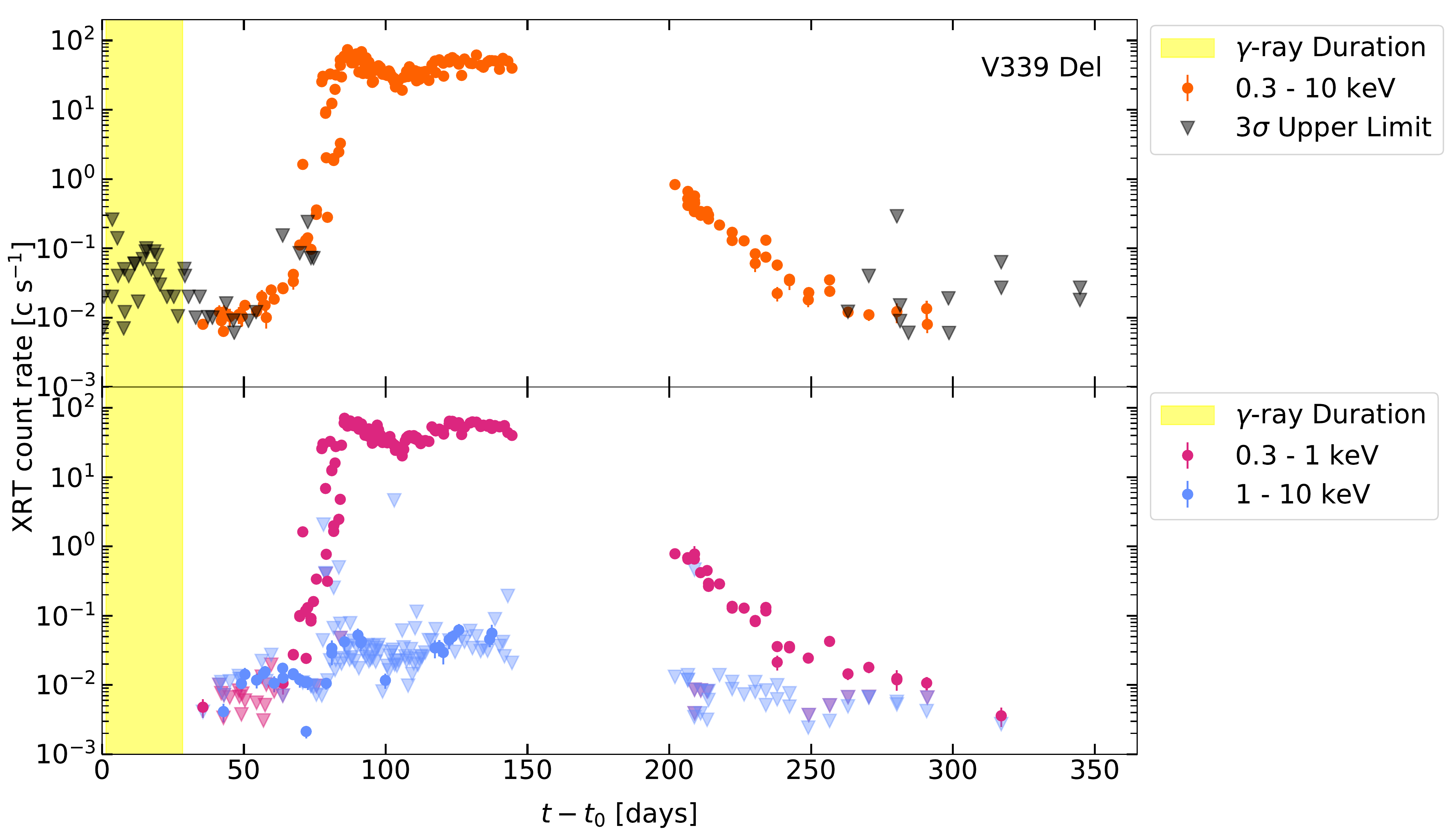}
    \caption{\textit{Swift}-XRT X-ray light curves of V339~Del. See Figure \ref{fig:V392_Per} for more details.}
    \label{fig:V339_Del}
\end{figure*}

\begin{figure*}[!t]
    \centering
    \includegraphics[width=0.8\textwidth]{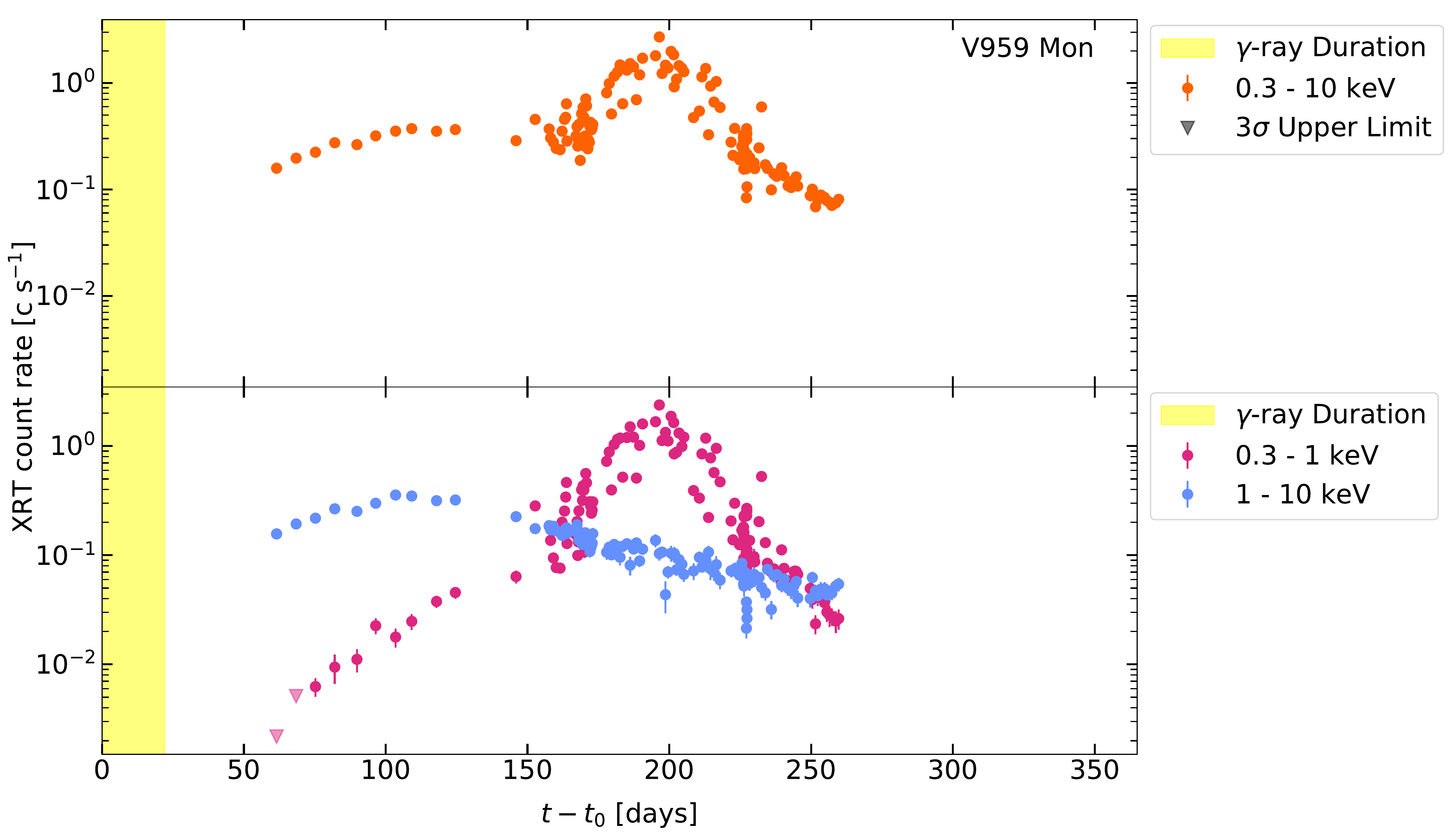}
    \caption{\textit{Swift}-XRT X-ray light curves of V959~Mon. See Figure \ref{fig:V392_Per} for more details.}
    \label{fig:V959_Mon}
\end{figure*}

\begin{figure*}[!b]
    \centering
    \includegraphics[width=0.8\textwidth]{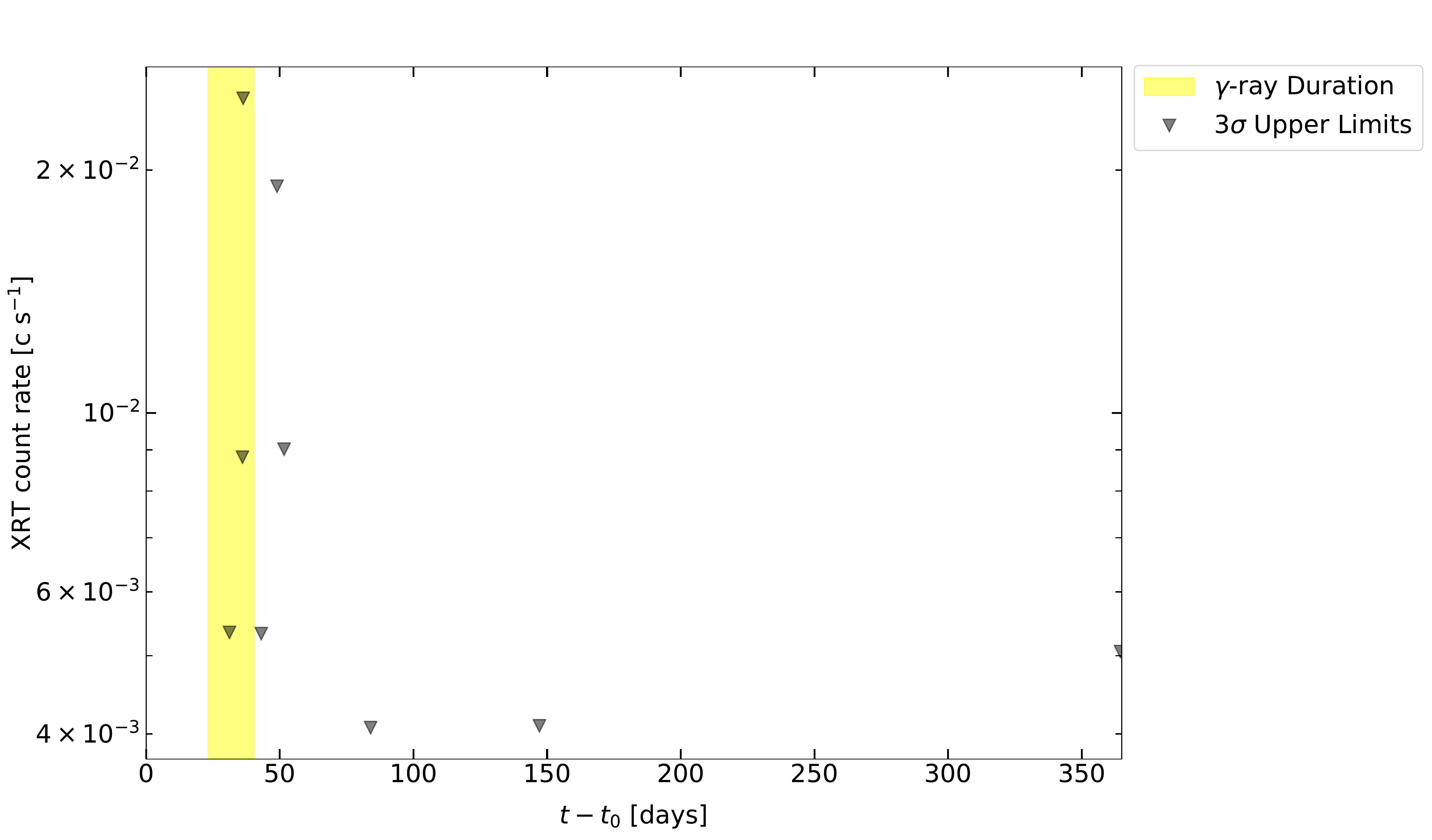}
    \caption{\textit{Swift}-XRT X-ray light curve of V1324~Sco. See Figure \ref{fig:V392_Per} for more details.}
    
    \label{fig:V1324_Sco}
\end{figure*}

\begin{figure*}[!t]
    \centering
    \includegraphics[width=0.8\textwidth]{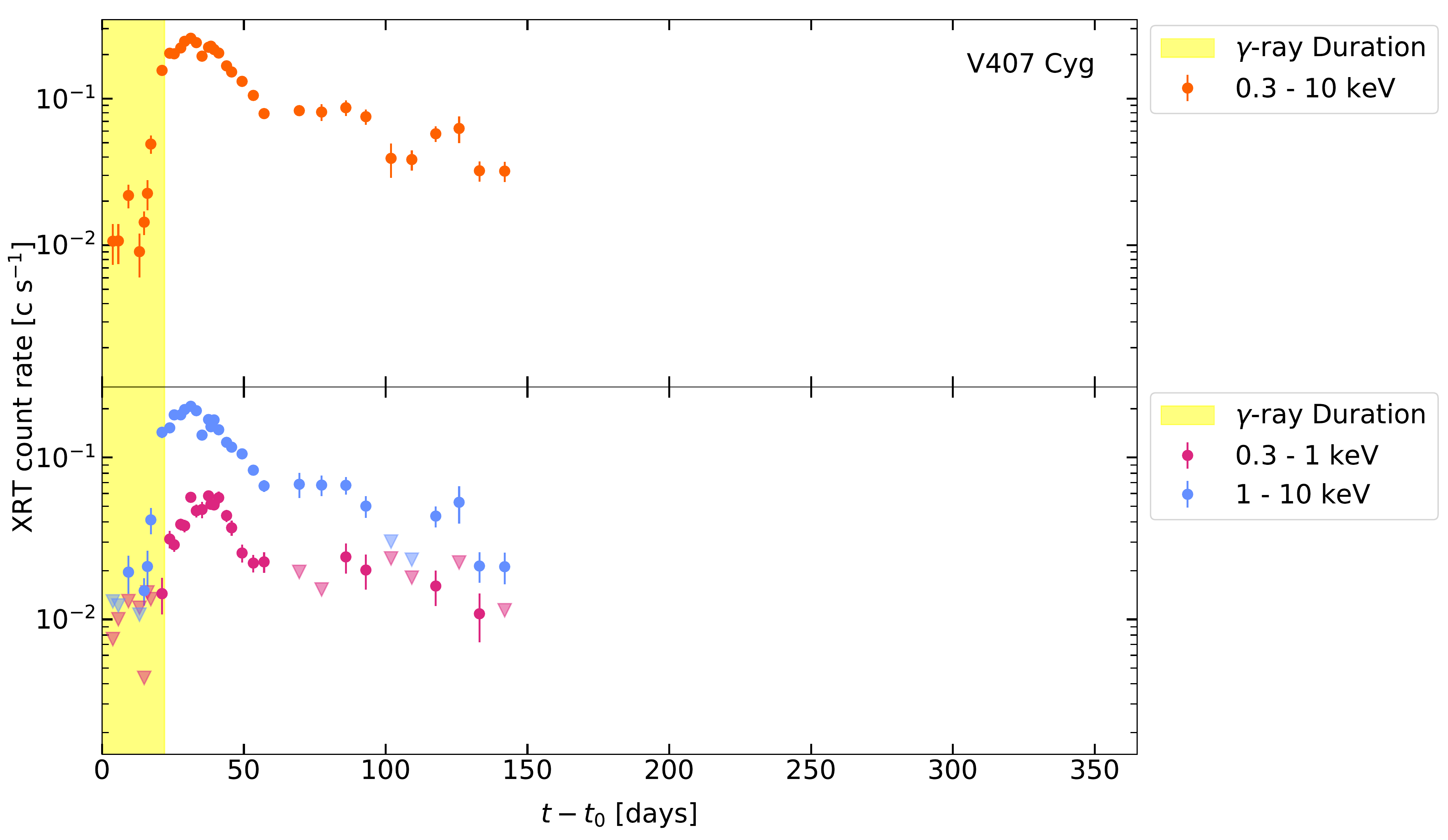}
    \caption{\textit{Swift}-XRT X-ray light curves of V407~Cyg. See Figure \ref{fig:V392_Per} for more details.}
    \label{fig:V407_Cyg}
\end{figure*}

\end{document}